\documentclass[preprint,12pt]{elsarticle}

\usepackage{amssymb}
\usepackage{amsmath}
\usepackage{siunitx}
\DeclareSIUnit{\wtpercent}{wt.\%}
\DeclareSIUnit{\volpercent}{vol.\%}
\DeclareSIUnit{\pixel}{px}
\usepackage{titletoc}
\usepackage{booktabs}
\usepackage{makecell}
\usepackage{multirow}
\usepackage[flushleft]{threeparttable}
\usepackage{float}
\usepackage{enumitem}
\usepackage{textgreek}

\usepackage{lineno}

\usepackage[editing]{coop-writing}
\cwnamedef{sabe}{red}{Santi}
\cwnamedef{lb}{blue}{Louis}
\cwnamedef{nj}{green}{Niels}

\journal{Acta Materialia}

\begin{document}

\begin{frontmatter}



\title{Iterative Thermodynamic Augmentation of Spatially Resolved Analytic Microscopy for Fast-Diffusing Solutes}


\author[LWT]{Santiago Benito} \ead{santiago.benito@rub.de}
\author[LWT]{Louis Becker}
\author[GEO]{Niels Jöns}
\author[LWT]{Sebastian Weber}

\affiliation[LWT]{organization={Chair for Materials Technology, Institute for Materials, Ruhr University Bochum},
            addressline={Universitätstr. 150}, 
            city={Bochum},
            postcode={44801}, 
            state={NRW},
            country={Germany}}
            
\affiliation[GEO]{organization={Institute of Geosciences, Ruhr University Bochum},
            addressline={Universitätstr. 150}, 
            city={Bochum},
            postcode={44801}, 
            state={NRW},
            country={Germany}}

\begin{abstract}
The spatially resolved quantification of fast-diffusing solutes presents several challenges in analytic microscopy. Given the critical role of interstitially alloyed elements in physical metallurgy, we propose a computational framework that addresses this limitation by augmenting spatially resolved composition maps of substitutional elements with computationally derived interstitial distributions. The underlying methodology is an iterative thermodynamic model: exploiting the stark differences in solid-state diffusion kinetics, the model assumes a state of partial chemical equilibrium exclusively for the mobile interstitial species. An optimization scheme iteratively adjusts a uniform interstitial chemical potential across the mapped microstructure until the integrated local concentration converges with an independently measured bulk value. Ultimately, this approach extracts thermodynamically consistent interstitial concentration maps from robust, low-noise microscopy data, yielding quantitative spatial arrays that are otherwise time- and resource-intensive to obtain at best.
\end{abstract}

\begin{graphicalabstract}
\includegraphics{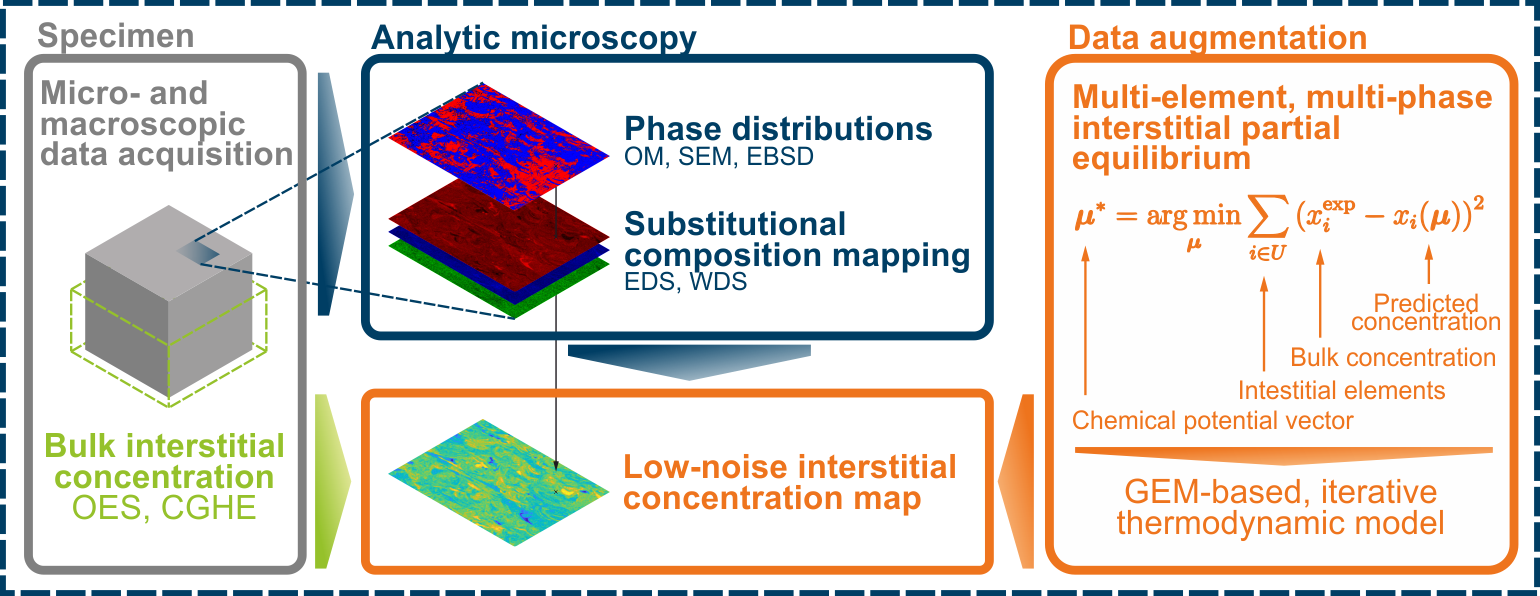}
\end{graphicalabstract}

\begin{highlights}
\item An iterative thermodynamic framework augments 2D and 3D analytic microscopy data
\item The method produces spatial distributions of interstitially alloyed fast-diffusers
\item The model assumes partial chemical equilibrium of the fast-diffusing species
\item Computed concentration maps are validated against WDS data
\item Use cases include multi-phase and synthetic microstructures
\end{highlights}


\end{frontmatter}



\section{Introduction}
\label{introduction}

Three core premises motivate this work:
\begin{enumerate}
    \item Interstitial diffusers influence the properties of metallic alloys considerably, even at low concentrations,
    \item Low concentrations of these light atoms are difficult to measure experimentally at best, and
    \item At sufficient homologous temperatures, their diffusion is  fast enough to enable us to make some simplifications toward equilibrium modeling.
\end{enumerate}

The positions that these atoms occupy in the host lattice and their electronic contribution dictate their impact as alloying elements. The mismatch between the available space within the lattice and the interstitial atom size induces localized lattice distortions \cite{Dieter.1988}. These, in turn, govern extended defect interactions, including dislocation pinning \cite{Hull.2011} and twin boundary mobility \cite{Christian.1995}. Parallel to the structural effects, the addition of electrons from these interstitial atoms directly alters the magnetic and electronic properties of the system \cite{Pepperhoff.2001}. Thermodynamic, mechanical, thermophysical, and magnetic properties all shift as a result, even at comparatively small concentrations.

Ferrous materials arguably offer the most common examples of this behavior. Carbon enhances the mechanical properties of steels through solid-solution strengthening. But, critically, the stark difference in carbon solubility in \textalpha- and \textgamma-iron leads to the formation of a variety of microconstituents---such as pearlite, bainite or martensite---each presenting distinct properties \cite{Krauss.2015}. Nitrogen exhibits analogous behavior while simultaneously improving localized corrosion resistance \cite{Gavriljuk.1999}.

Sensitivity to interstitial alloying extends beyond Fe-based systems. Oxygen and nitrogen dictate phase equilibria and mechanical properties in titanium and zirconium alloys \cite{Lutjering.2007}. Trace amounts of oxygen, nitrogen, or carbon govern the ductile-to-brittle transition temperature in refractory metals \cite{Tietz.1965}, and drive severe lattice distortion and phase stability in modern complex refractory alloys \cite{Zhao.2025}. Interstitial first generation, face-centered cubic high-entropy alloys similarly demonstrate that minute additions alter stacking fault energy and trigger strengthening \cite{Li.2017}.

Moving onto the second item on the list, the quantitative detection of interstitially alloyed elements is notoriously challenging due to the necessity of high spatial resolution analysis of low atomic number elements. Wavelength-dispersive electron beam microanalysis is the method of choice, because it can be done in-situ and with micrometer-scale resolution. However, light element analysis is challenging due to the low-energy of emission lines (typically below 1\,kV) the low fluorescence yield and the need for synthetic Bragg-diffraction pseudocrystals with large lattice spacing (i.\,e.~Layered Synthetic Microstructures, LSM). In addition, significant absorption of the soft X-rays can occur (need for large mass absorption coefficients) and the X-ray lines might show significant peak-shape effects in combination with severe interferences with L, M and N X-ray lines from heavier elements and higher diffraction order lines. Furthermore, light element analysis might suffer from ion migration, leading to time-dependant intensity changes (e.\,g.~\cite{Bastin.1988, Bastin.1990, Llovet.2021, Schweizer.2022}).

\begin{figure}[ht]
\centering
\includegraphics[width=0.7\textwidth]{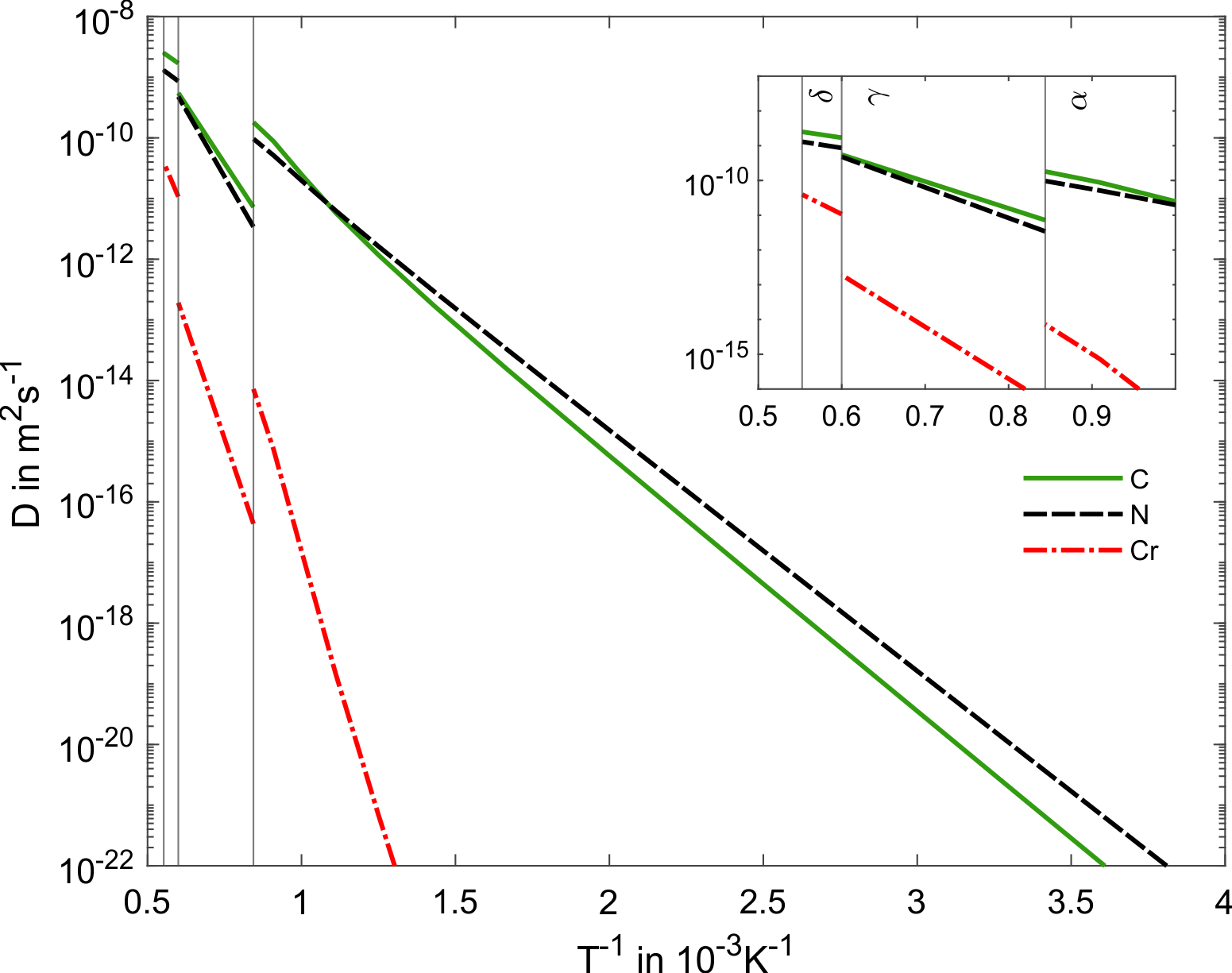}
\caption{Diffusivities of C, N, and Cr in solid Fe, extracted from the diffusion module of Thermo-Calc 2026a, using the databases TCFE15 and MOBFE9.}\label{fig:diffusion_coefficients}
\end{figure}

Figure~\ref{fig:diffusion_coefficients} illustrates the third and last point on the list. It shows, exemplarily, the diffusivities of C, N, and Cr in Fe as a function of temperature. These were computed with the Diffusion Module of the Thermo-Calc software 2026a and the databases TCFE15 and MOBFE9 for austenite and \textdelta- and \textalpha-ferrite. In the chosen representation, the diffusivities in each phase look like straight lines: Barring a slight curvature in the diffusivity of C, the coefficients conform to Arrhenius' exponential equation \cite{Balogh.2014}.

Nonetheless, the intended take-away message from Figure~\ref{fig:diffusion_coefficients} is the number of orders of magnitude that separate the diffusivities of C and N from that of Cr, particularly at lower temperatures. Interstitial diffusers, such as H, C, N, and O, are small atoms that migrate through the interstitial sites of the host lattice. Unlike in the diffusion of substitutional solutes, interstitial atoms do not require the formation of additional defects, such as vacancies. As such, they can move rapidly due to the low migration barriers between interstitial sites. This results in these high diffusion rates, which, in some cases, can be comparable to those in liquids \cite{Balogh.2014}. The size of the diffusing atoms, the distance between atoms in the lattice, and the site geometry (tetra- or octahedral) all influence the diffusion process.

The working hypothesis can thus be explicitly stated as follows:
\begin{quote}
\emph{In technological alloys, processing inevitably leads to the chemical segregation of heavy, substitutionally alloyed elements. This localized variance establishes a heterogeneous thermodynamic landscape across the microstructure. Because interstitial elements remain mobile at comparatively low temperatures, they redistribute to equalize their chemical potential, thereby achieving a state of spatial equilibrium.}
\end{quote}

However, any framework built upon this hypothesis carries two primary limitations: (i) the resolution of the modeled interstitial distribution is bound by the spatial resolution of the underlying analytic microscopy, and (ii) the localized segregation of interstitials to structural defects, such as grain boundaries or dislocations, cannot be captured. These boundaries define the model's operational domain, in principle, within the mesoscale. Because the underlying algorithm solely relies on compositional data, the source of the data is only contextually important. Regardless of whether the data on the substitutionally alloyed elements stems from energy- or wavelength-dispersive X-ray spectrometry, micro-X-ray fluorescence \cite{Yang.2024}, or is of synthetic origin, the data-agnostic nature of the approach will always produce thermodynamically consistent spatial distributions of interstitially alloyed elements. Furthermore, by isolating the pure, bulk thermodynamic component, which drives the distribution of the highly mobile species, from solute-defect interactions, the proposed method provides a solid foundation that can be built upon in the future. Both of these points will be discussed over the course of this work, particularly in the final subsection dealing with the outlook.

\begin{figure}[ht]
\centering
\includegraphics[width=\textwidth]{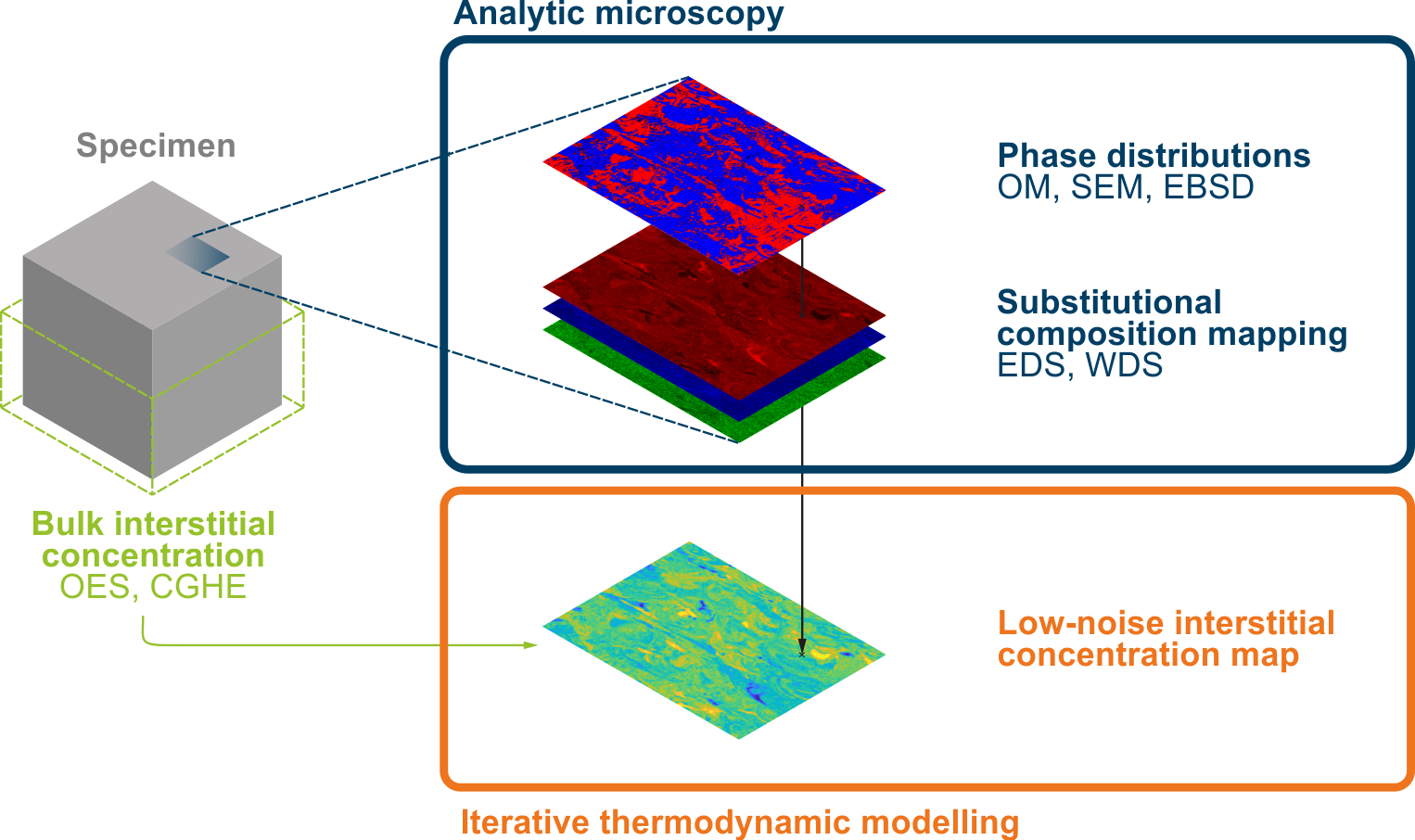}
\caption{Iterative thermodynamic augmentation strategy. By combining co-located phase and substitutional composition maps with macroscopic bulk measurements, an underlying partial chemical equilibrium model computes robust, spatially resolved concentration maps of fast-diffusing solutes. OM: Optical Microscopy; SEM: Scanning Electron Microscopy; EBSD: Electron Backscatter Diffraction; : Energy-Dispersive Spectrometry; WDS: Wavelength-Dispersive Spectrometry; OES: Optical Emission Spectrometry; CGHE: Carrier Gas Hot Extraction.}\label{fig:approach}
\end{figure}

Motivated by the high technological and scientific relevance of interstitial alloying, compounded with the challenges of its quantitative assessment, we present a robust method to estimate the spatial concentrations of these fast-diffusing elements in two- and three-dimensional fields. We propose employing Calphad calculations to augment composition maps obtained by more straightforward, low-noise measurements. Figure~\ref{fig:approach} schematically presents the strategy, which is described in detail in Section \ref{model}. It consists of an iterative thermodynamic model built upon serialized thermodynamic calculations and experimentally obtained composition maps and bulk interstitial concentrations. Section \ref{examples} presents a series of application examples---these demonstrate useful use-cases and validate the working assumptions and hypotheses. Finally, Section \ref{conclusion} provides a summary and an outlook.

\section{Iterative thermodynamic modelling}\label{model}

Iterative thermodynamic models encompass an overarching optimization strategy with embedded thermodynamic calculations within. They are not, in fact, by any means new. For instance, paraequilibrium calculations could serve as one example. Two phases are said to exist in paraequilibrium when the following three conditions are met on both sides of their interface at a given temperature and pressure \cite{Hillert.2012}:
\begin{enumerate}
    \item Equal chemical potential of fast diffusing elements, 
    \item Same ratio of the non-mobile alloying elements to the base element, and
    \item Equal combined weighted chemical potentials of the other elements.
\end{enumerate}

In other words, under paraequilibrium conditions, we can model phase transformations in systems with elements with very different diffusion coefficients. These transformations can happen faster than in complete equilibrium at the interface, sometimes referred to as orthoequilibrium \cite{Hillert.2004}.

In their standard mathematical representation, conditions 1-3 above can be written like so:

\begin{subequations} \label{paraequilibrium}
\begin{align}
\mu^j_i &= \mu^k_i, \forall i \in U, \\ 
x^j_s &= x^k_s, \forall s \in V, \\
\sum_{s \in V} \mu^j_s u^j_s &= \sum_{s \in V} \mu^k_s u^k_s,
\end{align}
\end{subequations}

where $j$ and $k$ are the phases being considered (such that $j \neq k$), $\mu$ is the chemical potential, $x$ is the mole fraction, and $u$ is the u-fraction defined as $x_s/(\sum_{k \in V} x_k)$. In Equations \ref{paraequilibrium}, the subscripts refer to elements, with $U$ being the set of all fast-diffusing elements and $V$ being the set of non-mobile elements in the system.

The optimization problem posed boils down to finding the chemical potential of the fast diffusers, $\boldsymbol{\mu} = \left( \mu_i \right)_{i \in U}$, such that the conditions in Equations (\ref{paraequilibrium}b) and (\ref{paraequilibrium}c) hold. For instance, the Thermo-Calc Toolbox for MATLAB\textsuperscript{\textregistered} V7 writes this problem like so \cite{TC-MATLAB.2014}:

\begin{equation} \label{paraeq-matlab}
\boldsymbol{\mu}^*= \underset{\boldsymbol{\mu}}{\arg \min} \left( \frac{\sum_{s \in V} \mu^j_s(\boldsymbol{\mu}) u^j_s(\boldsymbol{\mu})}{\sum_{s \in V} \mu^k_s(\boldsymbol{\mu}) u^k_s(\boldsymbol{\mu})} - 1 \right)^2,
\end{equation}

where the numerator and denominator result from equilibrium calculations from the phases $j$ and $k$, respectively. In the TC-MATLAB\textsuperscript{\textregistered} Toolbox, these thermodynamic calculations are solved through the Gibbs Energy Minimization (GEM) technique accessed through the Thermo-Calc API.

GEM is, in itself, an iterative algorithm that seeks for a minimum in the Gibbs free energy of the system. Originally proposed by Hillert, rather than relying on reaction pathways, the method considers, at a given temperature and pressure, all possible species simultaneously and identifies the set of phases and compositions at which no further decrease in Gibbs energy is possible \cite{Hillert.1981, Lukas.2011}. The method is at the heart of all Calphad-based software and, as such, is employed, in one way or another, in all major computational thermodynamics solutions, whether open-source or closed-source: Thermo-Calc, FactSage, pMELTS, openCalphad, etc \cite{ThermoCalcDoc.2025,FactSageDoc.2020,Ghiorso.2002,OpenCalphadDoc.2019}.

While a thorough description of the principles of the Calphad method is beyond the scope of this work, it is worth mentioning that most implementations enable the calculation of equilibria based on any set of boundary conditions---not just temperature, pressure, and system composition. Going back to the paraequilibrium example, the GEM algorithm can calculate the variables $\mu_s$ and $u_s$ in each phase with the following given conditions: present phase ($j$ or $k$, cf. Equations \ref{paraequilibrium}), temperature, pressure, u-fraction of the non-mobile elements, and activities of the fast-diffusing elements.

The explanation of the paraequilibrium calculation above is, as stated, meant as an introductory example. The following subsection provides an overview of current applications of iterative thermodynamic models in Materials Science and Engineering, as well as Mineralogy and Geophysics.

\subsection{Current applications of iterative thermodynamic models}
In Materials Science and Engineering, iterative thermodynamic models find applications in materials discovery and design, as well as the optimization of manufacturing processes. For example, Kusne et al. developed a closed-loop, autonomous system for materials exploration and optimization (CAMEO is the name they gave to it) to map phases and optimize properties of functional inorganic compounds, which led to the discovery of an epitaxial nanocomposite phase-change memory material \cite{Kusne.2020}. Sundar et al. applied a Calphad-based Bayesian optimization model to accelerate the discovery of high-temperature alloys, specifically exploring the nickel-chromium-cobalt-aluminum-iron composition space \cite{Sundar.2025}. Further, Liang et al. published an autonomous materials search engine (christened AMASE), which integrates experimental validation with Calphad calculations, to map the temperature-composition phase diagram of the Sn-Bi thin-film system \cite{Liang.2025}. In the same vein, Egels et al., Liang et al., and Howard et al. independently developed computational approaches that couple thermodynamic calculations with optimization strategies for application- and property-specific alloy design \cite{Egels.2020, Liang.2021, Howard.2024}. Other applications include the acceleration of Calphad calculations and the construction of phase diagrams themselves \cite{Shen.2025}. These applications, however, have an exploratory character and take the result of the underlying thermodynamic calculations at face value. In other words, they do not include specimen-specific data as inputs to refine, improve or extend the calculations.

Closer to the subject of the present work, recent studies have also begun coupling specimen-specific spatial data with thermodynamic computations to understand localized phase transformations. In particular, Li et al. directly imported EBSD-derived microstructural maps into cellular automata frameworks to simulate the time-dependent evolution of chemically banded Q\&P stainless steels \cite{Li.2025}. By calculating local chemical potentials via GEM to drive carbon diffusion and interface migration, their kinetic models demonstrated that assuming local equilibrium with negligible partitioning (LENP) at the interfaces most accurately captures the microstructural evolution of these highly segregated metastable states.

In this context, researchers in the geological sciences have produced software that is most similar to what we set out to achieve in this work regarding the augmentation of analytic microscopy data. The work by Duesterhoeft, Lanari, and Hermann on the open-source software XMap-Tools is of particular interest \cite{Duesterhoeft.2020, Lanari.2021}. While the software is a package for processing and visualizing compositional maps, offering functionalities like data classification, filtering, and statistical analysis, its key innovation is the direct integration of quantitative compositional maps of minerals, typically acquired via electron probe microanalysis (EPMA), as inputs for the model. The measured local bulk composition from a specific point in a rock sample is used to constrain GEM-based equilibrium calculations. In one of the examples the authors provided, the purpose of this integration is to find the unique pressure-temperature (P-T) conditions at which the thermodynamically predicted mineral assemblage, compositions, and proportions most closely match the actual measured data from that point in the rock. This iterative process of minimizing the misfit between prediction and observation enables a highly accurate reconstruction of a rock's specific P-T history, moving beyond blanket phase diagrams \cite{Lanari.2021}.

While these computational combinations extend the explanatory power of the retrieved data---whether to simulate time-dependent kinetic phase evolution or reconstruct a rock's P-T history---they do not enhance the compositional data itself. As previously detailed in the introduction to this work, the goal here is to generate thermodynamically consistent composition maps that include interstitial solutes from robust distribution information on the substitutionals, completely independent of a kinetic solver.

\subsection{Calculating the partial equilibrium distribution of fast diffusers}
Unlike in paraequilibrium calculations, the proposed model employs an independently measured concentration of the fast diffusers to establish the partial equilibrium. Mathematically, for composition maps $u(a,b)$ at the spatial locations $(a,b)$, we can write the conditions as follows \cite{Speer.2005, Hillert.2005}:

\begin{subequations} \label{partialequilibrium}
\begin{align}
\mu(a,b)_i &=  \mu_i, \forall i \in U, \\ 
\frac{1}{n}\sum_{a, b} x(a,b)_i &= x^\text{exp}_i, \forall i \in U.
\end{align}
\end{subequations}

Equation (\ref{partialequilibrium}a) states that the chemical potential of the interstitially alloyed elements in the system is independent of the coordinates $(a,b)$ and takes, for each $i \in U$, a constant scalar value. The second line in the equation system introduces the independently determined concentration of the fast diffusers, $x^\text{exp}$. Here, the mean mole fraction of each element $i$ across the $n$ positions of the composition map ne to equal the corresponding value $x^\text{exp}_i$. This description of the partial equilibrium is consistent with what other authors describe as \emph{constrained carbon equilibrium} (CCE) \cite{Traka.2024}, only is here generalized to all interstitially alloyed elements.

Echoing Equation \ref{paraeq-matlab}, the optimization problem can be defined as:

\begin{equation} \label{partialeq-optimization}
\boldsymbol{\mu}^*= \underset{\boldsymbol{\mu}}{\arg \min} \sum_{i \in U} \left( x_i^{\exp} - x_i(\boldsymbol{\mu}) \right)^2
\end{equation}

Equations \ref{partialequilibrium} and \ref{partialeq-optimization} do not explicitly refer to the phases, as solute partitioning is implicitly accounted for through the map coordinates. To put it differently, different regions in the composition maps can be host to different phases.

\begin{figure}[p]
\makebox[\textwidth][c]{\includegraphics[width=1.35\textwidth]{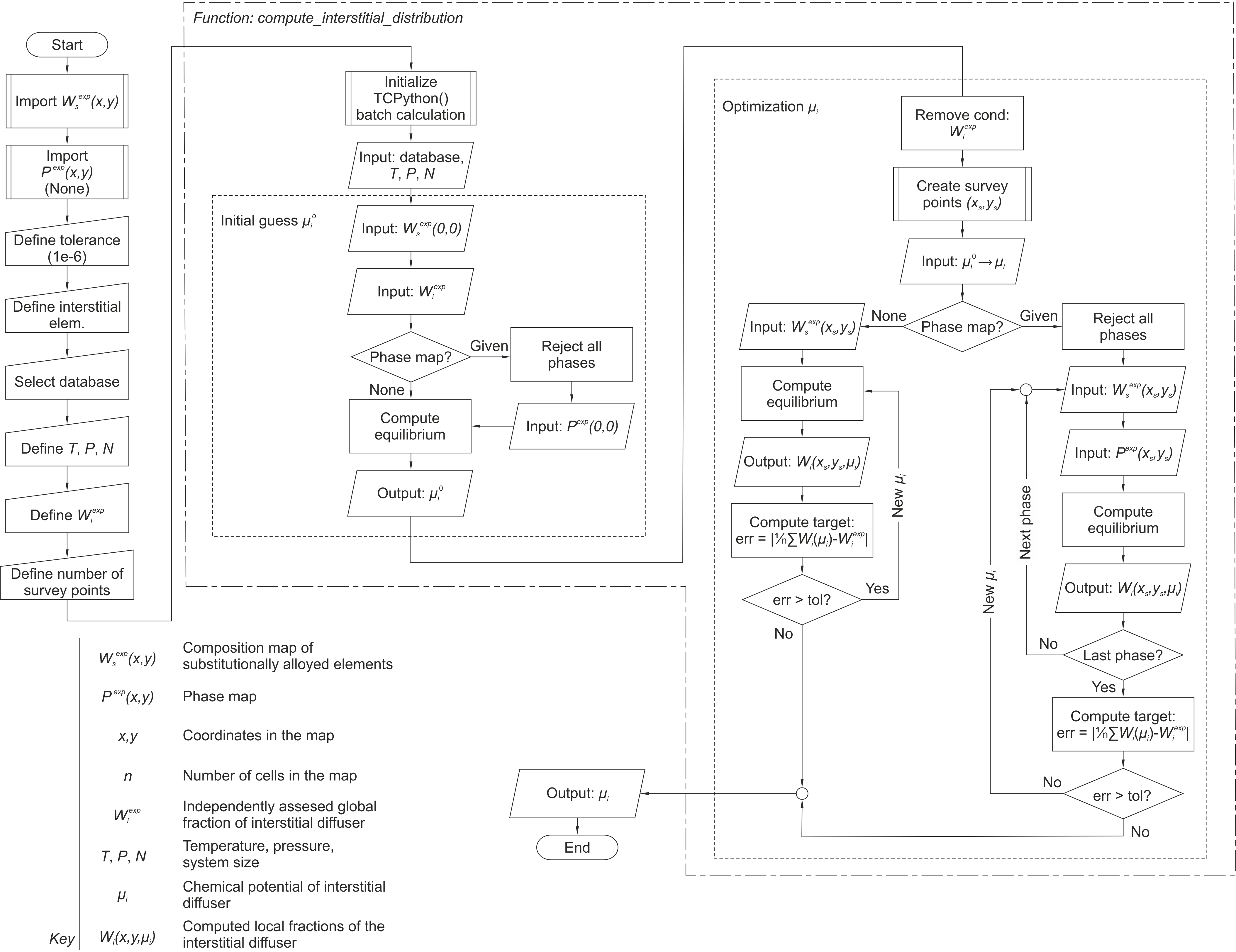}}
\caption{Flowchart of the proposed algorithm. See the embedded key for reference of the employed nomenclature.}\label{fig:flowchart}
\end{figure}

Figure~\ref{fig:flowchart} presents the flowchart of the solution employed to solve the optimization problem posed above. The implementation was carried out in Python using the TC-Python Software Development Kit (SDK) developed by Thermo-Calc Software \cite{TCPython.2025}. Using the provided application programming interface (API) and PyCharm Community 2021.3.2, the programmed function \texttt{compute\_interstitial\_distribution} performed the required equilibrium calculations in Thermo-Calc 2026a employing the TCFE15 database.

The solution starts with the import of the composition maps and, optionally, the phase map, followed by the usual system and database definitions befitting any Calphad calculation. Critically, and as opportunely mentioned, the input of the independently, experimentally determined amount of the interstitial elements is also to be provided. The program generates an initial guess of the chemical potential of each considered interstitially alloyed element by selecting a random point in the field and setting its corresponding mole fraction to the experimentally measured value.

In scanning electron microscopy, provided that the required detector is fitted to the microscope, composition maps can be recorded concurrently with spatially resolved electron backscattered diffraction (EBSD) data. Among the possibilities of EBSD is the creation of the corresponding phase maps. These provide complementary information on which phase is present at every position in the field of view, as long as the location can be correctly indexed from the electron diffraction patterns. Phase maps can also be defined based on prior knowledge and context, such as assuming a single phase across the entire domain based on predicted phase dissolution temperatures, or by extracting distinct regions from the element distribution maps themselves using compositional segmentation tools. If the function \texttt{compute\_\allowbreak interstitial\_\allowbreak distribution} is fed a phase map along with the composition maps, the calculation of each position $(a,b)$ is instructed to reject all phases and reinstate only the corresponding, present ones, and the global optimization of the GEM algorithm is deactivated. Otherwise, the calculation finds the global minimum in the Gibbs energy landscape, as expected. This course of action is taken for both the setup of the chemical potential initial guesses and for the optimization itself. 

Since composition maps can easily require more than a million equilibrium calculations (a medium-sized $1000 \times 1000$ px map can indeed have that many unique compositions), a series of operational improvements was implemented to reduce the time required to run the optimization. These include using batch equilibrium calculations and parallelization methods from \texttt{concurrent.futures} \cite{BrianQuinlan.2025}. Working with survey points provides a key speed increase, too. This approach randomly selects a pre-defined number of points $(a,b)$ in the map (dubbed $(a_s,b_s)$ in Figure~\ref{fig:flowchart}) and uses those as a representative sample of the whole map to run the calculations.

Some applications require ignoring some map locations. Such would be the case, for instance, when dealing with porosity or other microstructural defects, such as non-metallic inclusions, or even constituents that can---or should---be safely set aside for the interstitial distribution calculations. This feature is made available by labelling these regions in the corresponding compositional and phase maps as \texttt{NaN}, as the program ignores these positions.

The function \texttt{fmin} in the Python module \texttt{scipy\_optimize},  a downhill simplex algorithm, conducts the optimization proper \cite{Virtanen.2020}. The algorithm will minimize the optimization function described in Equation \ref{partialeq-optimization} until the absolute difference between the best and worst function values in the current simplex is less than the parameter \texttt{ftol} \emph{and} the maximum difference between the vertex coordinates is less than the parameter \texttt{xtol}. The parameters \texttt{xtol} and \texttt{ftol}, along with all other optional arguments of the Python implementation of the Nelder-Mead algorithm, can be adjusted as needed.

\section{Application examples}\label{examples}

\subsection{One-dimensional example} \label{one-dimensional-example}
This work addresses two- and three-dimensional fields; this first application example, however, is one-dimensional. The intent behind this choice is twofold: Firstly, we propose comparing the results of the proposed algorithm with those obtained from selected one-dimensional diffusion simulations. Secondly, we show how this approach can replace such simulations in the future.

Starting with the second point in the previous paragraph, let us first point out some works that relied on finite element-based solvers to approximate fast-diffuser partial equilibrium conditions from one-dimensional substitutional element distribution data. Mujica Roncery et al., for example, employed EDS line scans describing the variability of Mn and Cr and one-second DICTRA simulations to estimate the carbon and nitrogen distribution in a novel Fe-Mn-Cr-C-N TWIP steel \cite{MujicaRoncery.2010}. Becker et al., on the other hand, estimated the distribution of nitrogen in a hot isostatically pressed, diffusion-alloyed high-nitrogen stainless steel using a sample line extracted from an EDS composition map and a 60-second DICTRA simulation \cite{Becker.2023}. DICTRA is a diffusion simulation module based on Calphad data, currently commercialized as an add-on to Thermo-Calc. Its underlying engine is an FEA solver that solves the diffusion partial derivative equations in multi-component systems \cite{Borgenstam.2000}. Critically, DICTRA is currently limited to one-dimensional simulations. 

These works have in common that they were not really interested in the short-term kinetics of the system under scrutiny. Instead, these studies employed diffusion simulations under the assumption that the fast diffusers will \emph{redistribute} themselves from a constant-composition starting point, thereby approximating a state of local equilibrium with negligible partitioning \cite{Coates.1973, Li.2025}---albeit driven by pre-existing chemical inhomogeneities rather than true phase partitioning. The total simulated time was chosen arbitrarily and is not a variable of genuine interest. It is, rather, an instrument to approximate a partial equilibrium distribution, based on the fact that the interstitial elements have mobilities orders of magnitude higher than those of the substitutionally alloyed elements that create the chemical potential landscape in the first place. Therefore, the only limitation placed on the selection of the simulation time is not to allow these substitutional solutes to diffuse, thus completely covering the conditions of partial equilibrium described in Equations \ref{partialequilibrium}.

The one-dimensional example presented in the following stems from the microstructural analysis of a powder mixture consisting of the austenitic stainless steel DIN EN X2CrNi18-9 (within the composition range of AISI 304L, cf. Table \ref{304L_composition}) and Si\textsubscript{3}N\textsubscript{4}. The powders were mechanically mixed in a tumbling device with 0.9 mass percent silicon nitride and subsequently processed by laser powder bed fusion using the shell-core approach, in which a dense shell surrounds a still powdery core. For detailed information we refer to \cite{Becker.2024}. The subsequent hot isostatic pressing (HIP) at \SI{1200}{\degreeCelsius} and \SI{150}{\mega\pascal} for \SI{5}{\hour}, was followed by uniform rapid quenching (URQ\textsuperscript{\textregistered}). URQ\textsuperscript{\textregistered} is a proprietary high-rate cooling mode available in the HIP unit employed in this study, a Quintus AB Technologies model QIH 9. The approach is similar to previously published work on diffusion alloying to achieve high nitrogen concentrations in stainless steels \cite{Becker.2024}. Cr-Ni austenitic stainless steels otherwise have too low N solubilities in the liquid phase, rendering attempts to achieve the desired alloying in conventional manufacturing fruitless. The amount of silicon nitride to additivate was such that the maximum solubility of N in austenite was reached during the HIP process. The expectation is that after URQ\textsuperscript{\textregistered}, the microstructural state is frozen down to ambient temperature, i.e., any chemical segregations are fixed and no new phases form.

\begin{figure}[htb]
\makebox[\textwidth][c]{\includegraphics[width=0.8\textwidth]{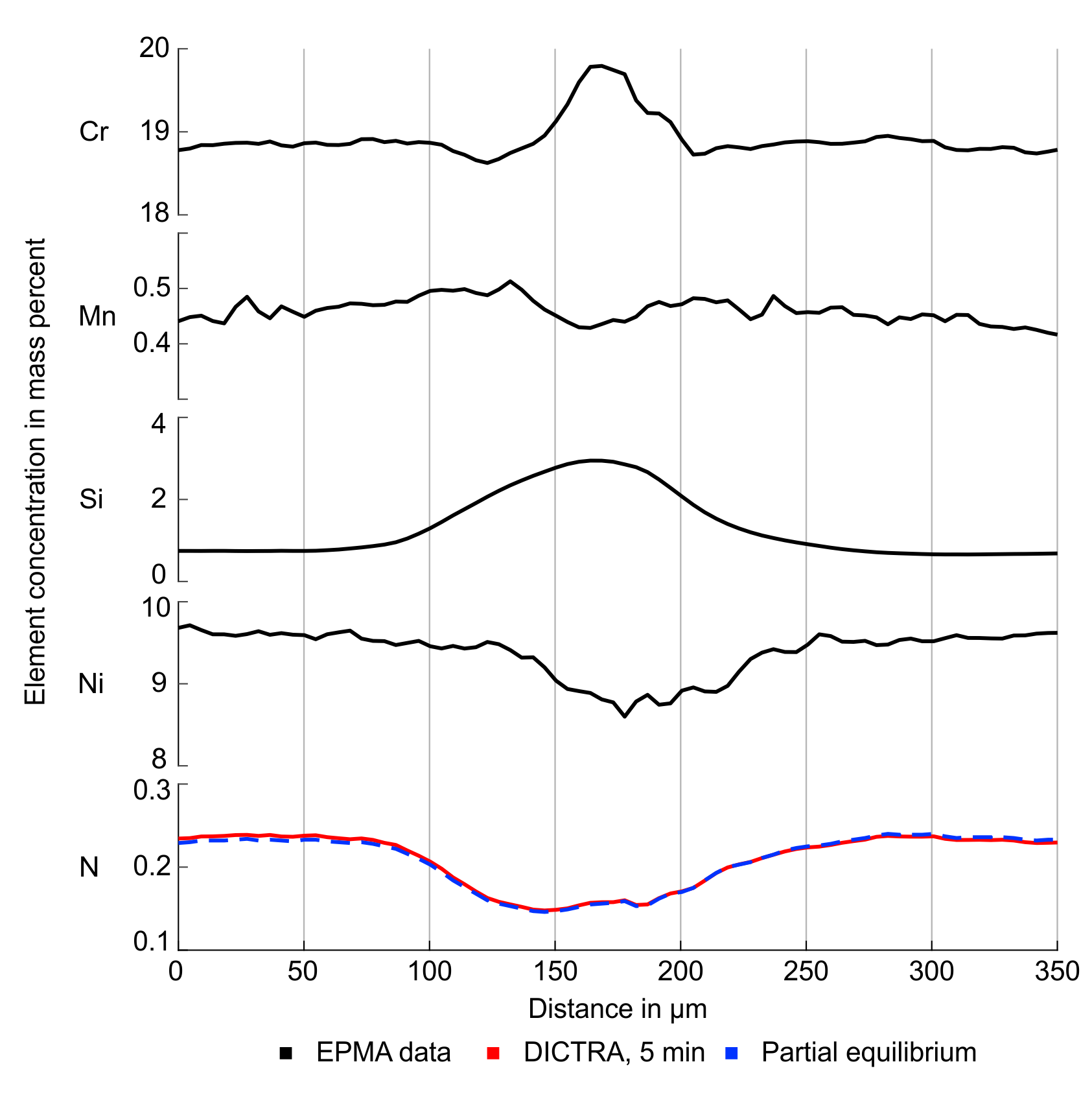}}
\caption{EPMA concentration line profiles of chromium, manganese, silicon, and nickel in the diffusion-alloyed austenitic stainless steel. These are plotted alongside nitrogen distributions computed via a short-term diffusion simulation and the proposed partial equilibrium algorithm.}\label{fig:1d-example}
\end{figure}

Figure~\ref{fig:1d-example} presents, from top to bottom, profiles of chromium, manganese, silicon, and nickel measured with a Cameca SX5FE field emission electron microprobe with its five wavelength dispersive spectrometers (WDS). The device was operated with a 10 kV acceleration voltage and 400 nA beam current. Quantification of the raw data followed the procedure of Merlet \cite{Merlet.1994}. The selected reference materials were pure metal samples for Fe, Mn, and Ni, whereas Cr\textsubscript{2}O\textsubscript{3} and CaSiO\textsubscript{3} were employed for Cr and Si, respectively. Where necessary, interference corrections for higher order X-ray lines were performed using empirically determined correction factors. The probe line starts in what originally was a steel powder, traverses a previous silicon nitride particle, and finally finishes back in the steel. This route is reflected in the increased silicon content with an onset around \SI{75}{\micro \metre} into the linescan. During the HIP process, chromium and iron diffused into the Si-rich zones, whereas the manganese and nickel alloyed in the steel showed a relative depletion in this range. After quenching to room temperature, these substitutional solutes can be considered to be in a metastable equilibrium \cite{Becker.2024}.

The bottom row in Figure~\ref{fig:1d-example} plots the results of the partial equilibrium calculation, along with the DICTRA simulation at a simulated time of five minutes. Both were conducted at the HIP conditions described above. To generate the nitrogen profiles, the compositional data shown in the previous rows, with iron as the basis element, were employed and the phase was fixed as austenite (\texttt{FCC\_A1}). The total nitrogen content in the system was \SI{0.213}{\wtpercent} and was independently determined through carrier gas hot extraction. Both approaches relied on Thermo-Calc 2026a and the database TCFE15, with the diffusion module additionally drawing mobility data from the database MOBFE9. 

As expected, the agreement between N-distributions is excellent: the selected simulation time seems, at a glance, to be enough to allow for a partial equilibrium redistribution of nitrogen. A more in-depth examination of the profiles, however, reveals that some minor differences exist. While the diffusion simulation starts out predicting slightly larger values (at the first data point, DICTRA returns \SI{0.2342}{\wtpercent} and the partial equilibrium calculation \SI{0.2274}{\wtpercent}), this trend reverses with increasing distance (at \SI{350}{\micro \metre} \SI{0.2299}{\wtpercent} and \SI{0.2317}{\wtpercent} for DICTRA and partial equilibrium, respectively). The source of these discrepancies is the selected simulation time. After five minutes at \SI{1200}{\degreeCelsius}, the system still exhibits a slight chemical potential gradient of nitrogen. Longer running eventually leads to partial equilibrium conditions but also to some degree of mobility in the substitutionally alloyed elements. Figure~\ref{fig:1d-example-dictra} presents the results of different simulation times in the diffusion module of Thermo-Calc.

Blame in the diffusion simulation would be misplaced. Co-opting Thermo-Calc's diffusion module to approximate any \emph{equilibrium} condition is in itself a tenable, yet arguably sub-optimal solution. The simulation starting point is not the system state at the beginning of the HIP consolidation process or at any other arbitrary time; rather, it is the end state. This means that any amount of simulation time will artificially extend the consideration of the kinetic system for exactly the selected simulation time. To summarize, since the segregation state prior to the high-pressure quenching step is impossible to obtain, there is no viable approach to solve the problem at hand with diffusion simulations without any movement of the substitutional solutes.

Arguably, the selected example is an extreme case of chemical segregation. The premise of diffusion alloying---especially when employing such different starting species---virtually guarantees strong chemical potential gradients. In systems presenting segregations associated with solidification processes of a single species, the diffusion simulations are expected to perform better. However, this system choice was a conscious decision to highlight the limitations of arbitrary time frame selection and, most importantly, of the diffusion-based approach itself.

The argument mounted above speaks for the need to replace short-term diffusion simulations with simpler and faster partial equilibrium calculations of the fast diffusers. The benefits of this proposition compound with their straightforward algorithmic extension to two and three dimensions. The following subsections present two-dimensional use-cases, including a validation with WDS nitrogen maps.

\subsection{Two-dimensional validation example} \label{two-dimensional-validation}

Similar to the diffusion alloying use-case shown in the one-dimensional example, we decided to carry out the two-dimensional demonstration and validation on a powder mixture. Instead of N-alloying into a DIN EN \mbox{X2CrNi18-9}, this example shows elemental maps of a mixture of Fe20Cr (cf. Table \ref{fe20cr_composition}) and Si\textsubscript{3}N\textsubscript{4}. The technical motivation behind this combination is the synthesis of manganese- and nickel-free austenitic stainless steels. The alloying concept is therefore based on austenite stabilization through nitrogen alloying. This concept has been previously demonstrated in Ref. \cite{Becker.2025}. The data presented here corresponds to an as-HIPed specimen. The addition of Si\textsubscript{3}N\textsubscript{4} was carried out at a \SI{2}{\wtpercent} mixing ratio in a tumbling device. This amount corresponds, analogous to the one-dimensional example, to the maximum expected solubility of N (approx. \SI{0.8}{\wtpercent}) at the processing conditions. The HIP parameters were kept identical as in the example above (\SI{1200}{\degreeCelsius}, \SI{150}{\mega\pascal}, \SI{5}{\hour}, URQ\textsuperscript{\textregistered}). 

To collect the element distribution, we employed the previously described Cameca SX5FE field emission electron microprobe. To avoid absorption effects by carbon contamination, the sample was properly cleaned within a plasma cleaner prior to analysis. The X-ray lines and diffraction crystals were: Cr K\textalpha (LLiF), Fe K\textalpha\ (LLif), Si K\textalpha\ (TAP), O K\textalpha\ (LPC0) and N K\textalpha\ (LPC2; see the supplementary data \ref{fe20cr_maps}). Pulse height analyser (PHA) settings in X-ray flow through counters were mostly operated in integral mode, except for oxygen and nitrogen measurements, where a differential mode was chosen. However, the differential window was kept fairly wide to avoid pulse height depression artefacts. The element distribution maps were acquired using a dwell time of \SI{0.045}{\second} per pixel at \SI{10}{\kilo\volt} acceleration voltage and \SI{170}{\nano\ampere} probe current. Using maps of the elements Cr and Si (with Fe as the basis element) and a global nitrogen content of \SI{0.8}{\wtpercent} (obtained through carrier gas hot extraction, CGHE), we computed the partial equilibrium N distribution with all points on the map fixed to the \texttt{FCC\_A1} phase. The experimentally determined N map was subsequently used as the ground truth to evaluate the performance of the iterative approach.

The N-distribution was exported as a raw counts map. In addition, fully quantitative spot measurements were acquired. Spot locations within the map area were chosen to cover the full elemental concentration range. Quantification of the raw data followed the procedure of Merlet \cite{Merlet.1994}; Cameca mass absorption coefficients were applied. Reference materials were metallic Fe for Fe, Cr\textsubscript{2}O\textsubscript{3} for Cr, CaSiO\textsubscript{3} for Si, MgO for O and BN for N. A beam current of \SI{145}{\nano\ampere} at \SI{10}{\kilo\volt} acceleration voltage was used. The quantitative results (in wt.\%) were used with a linear regression curve to quantify the raw intensity maps. Processing of the map consisted of median filtering with a box kernel of side \SI{3}{\pixel}, followed by a Gaussian filter with a size of \SI{7}{\pixel}. These operations targeted the noise reduction to emphasize the underlying microstructural features. Care was taken to avoid data loss: with a pixel scale of \SI{1}{\pixel\per\micro\metre} and a covered area of 1024\,$\times$\,\SI{768}{\micro\metre}, the filter sizes correspond to length scales below previously observed compositional gradients in these mixtures \cite{Becker.2025} while suppressing stochastic detector noise.

\begin{figure}[tb]
\makebox[\textwidth][c]{\includegraphics[width=\textwidth]{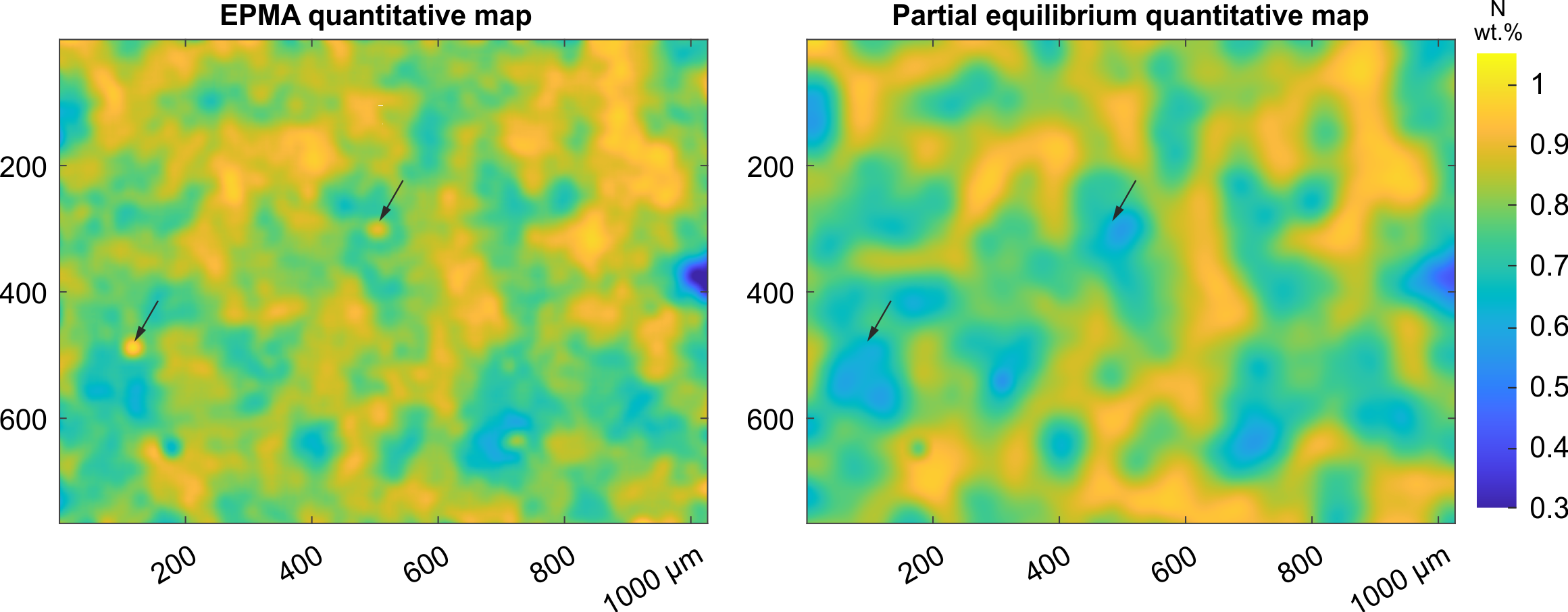}}
\caption{EPMA nitrogen concentration map and independently computed partial equilibrium via the proposed partial equilibrium algorithm.}\label{fig:2d-example-maps}
\end{figure}

Figure~\ref{fig:2d-example-maps} shows the WDS-detected spatial distribution of nitrogen in a side-by-side comparison with the result of the computation. Qualitatively, the agreement is good: regions with higher predicted nitrogen contents closely mirror those with higher measured nitrogen concentrations. The localized discrepancies---marked with arrows in both maps---likely correspond to nitrides. Since we assumed for this validation example a single austenitic phase for the entire domain, these experimentally observed high-nitrogen features are naturally absent in the partial equilibrium calculation. 

\begin{figure}[tb]
\makebox[\textwidth][c]{\includegraphics[width=\textwidth]{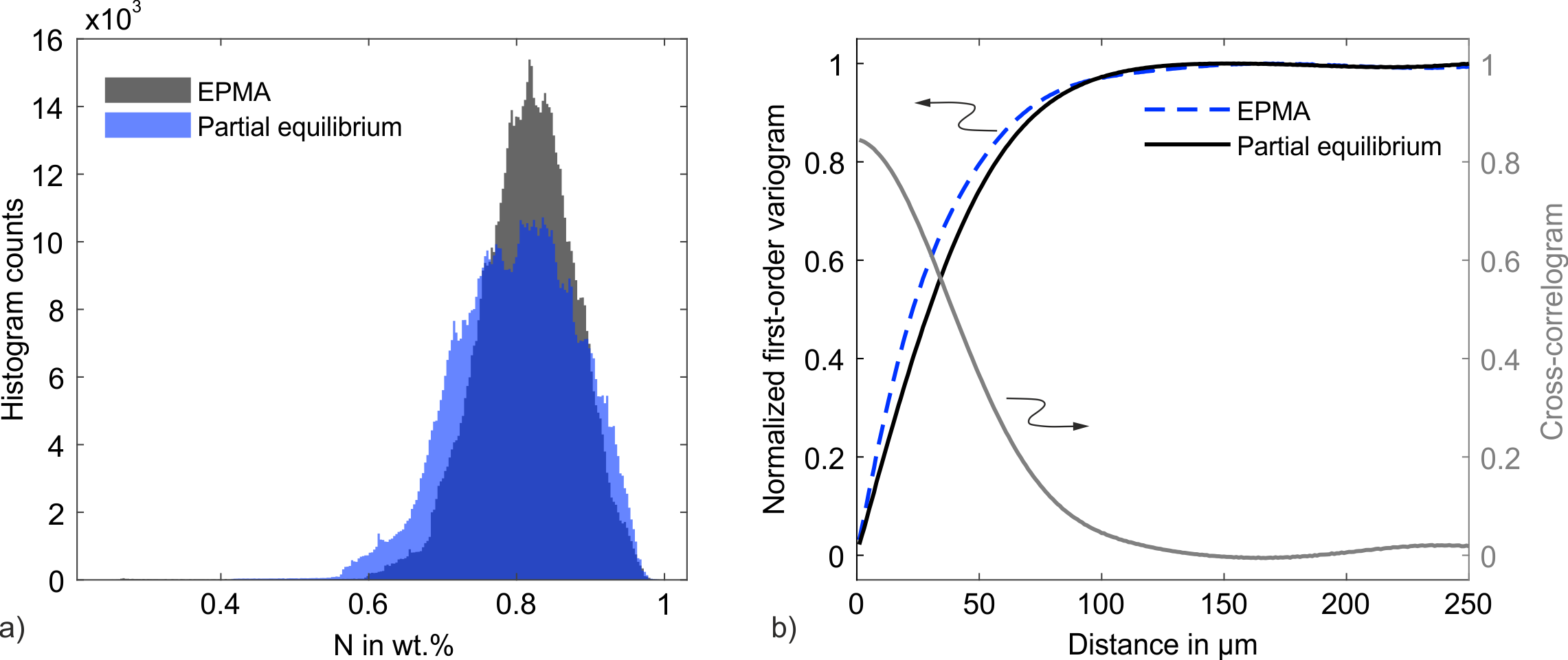}}
\caption{Global and spatial statistics of the EPMA and partial equilibrium nitrogen distribution data. (a) Histograms. (b) Radially averaged, normalized first-order variograms and radially averaged cross-correlogram between distributions.}\label{fig:2d-example-statistics}
\end{figure}

Beyond the qualitative assessment above, a quantitative comparison can be derived from the histograms and variograms of the maps. These are plotted in Figure~\ref{fig:2d-example-statistics} (a) and (b), respectively. Focusing on the histograms first, these exhibit a high degree of overlap. Both distribution shapes are similar and approximate normal distributions, albeit with a thicker left shoulder. The mean values come out to be \SI{0.81}{\wtpercent} and \SI{0.8}{\wtpercent} for the measured and computed N concentrations, respectively, while the standard deviations are, again, comparable at \SI{0.066}{\wtpercent} (EPMA) and \SI{0.081}{\wtpercent} (partial equilibrium). That the mean nitrogen content of the partial equilibrium distribution resulted in \SI{0.8}{\wtpercent} comes as no surprise; this is exactly the global value given as an input for the partial chemical potential homogenization. The mean content 0.81 wt.\% extracted from the WDS measurement does not perfectly correspond to the independently assessed global value, thus explaining that small modelling discrepancy.  Overall, the agreement on this aggregate level is very good.

It is worth mentioning that the spread of the histograms is dictated, beyond the microstructure itself, by both stochastic measurement noise and the map processing steps. As noted previously, the processing steps selected are those that highlight the structure without compromising the data. Due to the subjectivity of this matter, it is likely that a different team would settle on a different processing routine---regardless, we maintain that the present use-case accurately represents a typical workflow to obtain, prepare, and publish compositional maps.

The variograms complement these aggregate considerations. Originally a tool born from the need to characterize soils, they provide information on spatial correlations \cite{Benito.2023}. Figure~\ref{fig:2d-example-statistics} (b) presents the radially averaged first-order variograms for both the experimental and predicted distributions. To isolate the spatial correlation structure from the absolute variance discrepancies discussed previously, the curves are normalized to unity (left y-axis). The inter-map cross-correlogram is superimposed on the secondary y-axis. Interested readers can find the mathematical definitions of these spatial descriptors in the supplementary data (\ref{normalized-variogram} and \ref{cross-correlogram}); for more information, see Ref. \cite{Marcotte.1996} for the original implementation and Ref. \cite{Benito.2023} for their application on composition and property maps.

Physically, the variograms display the expected behavior for real-valued property maps \cite{Pachauri.2022}: the compositional dissimilarity is minimal at small separation distances (sometimes called lags) and rises monotonically with increasing spatial separation. The curves eventually plateau at a correlation range of approximately \SI{100}{\micro\metre}. This distance corresponds to the characteristic length scale of the segregation microstructure, likely reflecting the average spacing between Si\textsubscript{3}N\textsubscript{4} particles in the mixture. Critically, the normalized experimental and computed curves largely coincide throughout the entire range. The small deviations that can be observed at distances up to \SI{100}{\micro\metre} likely stem from the random measurement error and the map processing employed to mitigate it.

The agreement observed in the variograms demonstrates that the computation successfully predicts not only the bulk distribution of nitrogen but also its specific spatial arrangement. This claim can be further substantiated by considering the inter-map cross-correlogram. The cross-correlogram returns values between 1 (perfect correlation) and -1 (perfect anticorrelation), with 0 indicating no correlation at all. At zero distance, the value corresponds to the mean of the main diagonal of the correlation matrix, thus representing the Pearson correlation coefficient when the maps are perfectly aligned. Here, we obtained a value of approximately 0.85. While theoretically a value of unity is possible, due to the previously mentioned stochastic noise this is practically not attainable.

Finally, with increasing pixel-pair distance the cross-correlation drops, reaching a value of zero at approximately \SI{100}{\micro\metre}. The agreement with the structural information yielded by the variograms confirms that the reconstruction is quantitatively, again, within the stochastic error of the WDS system, identical to the measurement. In other words, the model generates a composition map that is indistinguishable from the measurement in terms of spatial statistics with the added advantage of being thermodynamically consistent with robust mappings of elements with higher atomic number, in this case chromium. For an additional correlation analysis, please see \ref{additional-correlation}.

\begin{figure}[tb]
\makebox[\textwidth][c]{\includegraphics[width=\textwidth]{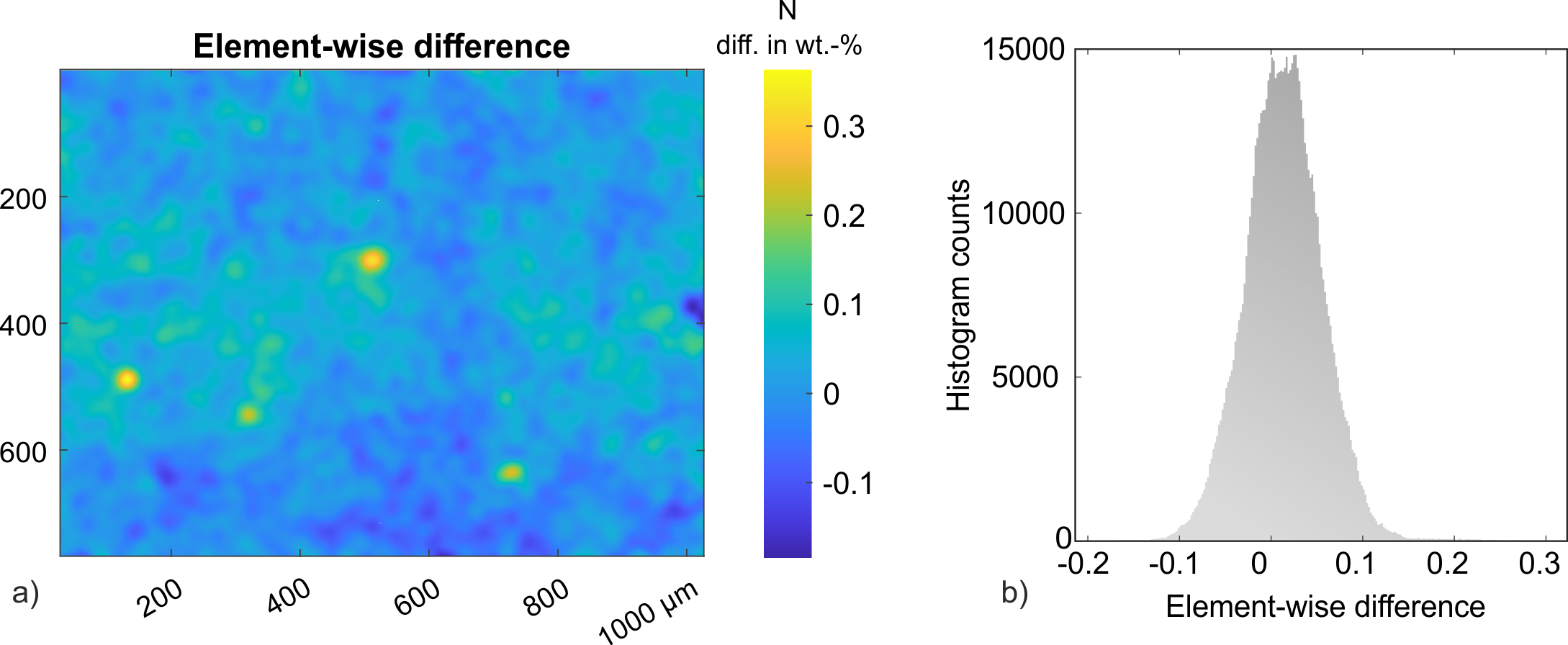}}
\caption{Element-wise difference between the EMPA and partial equilibrium distribution of nitrogen. (a) Spatial distribution. (b) Histogram.}\label{fig:2d-example-difference}
\end{figure}

To conclude the quantitative analysis, we will now consider the local, pixel-wise deviation between the calculation and the experiment. Figure~\ref{fig:2d-example-difference} (a) displays the spatial distribution of the residuals, i.e., the element-wise subtraction of the measured values from the predicted ones. Other than two bright spots, the maps lack distinct microstructural features. These circular shapes were opportunely discussed and likely are chromium nitrides not considered in the calculation. The featureless background indicates, once again, that the model has successfully predicted the systematic chemical variance, shedding the random background variance.

The histogram of the residuals in Figure~\ref{fig:2d-example-difference} (b) provides a quantiative perspective. The distribution shows a high degree of centralization, with 95\% of all pixels deviating by less than $\pm$ \SI{0.087}{\wtpercent} from the EPMA data. To contextualize this information and considering the global nitrogen content of \SI{0.8}{\wtpercent}, this confidence interval of approximately $2\sigma$ corresponds to a relative deviation of roughly 11\%.

Despite utilizing a high probe current of \SI{170}{\nano\ampere} and a dedicated synthetic multi-layer crystal (LPC2) to maximize counts, the mean nitrogen signal was recorded at an average of 45 counts per pixel (the dwell time was, again, \SI{45}{\milli\second}). According to the Poisson statistics of the data acquisition technique, the absolute intrinsic lower bound of uncertainty can be approximated as $\sigma \approx \sqrt{N}$, where $N$ is the number of counts \cite{Goldstein.2018}. Applying this relationship yields a theoretical floor of 15\%, a value well above the observed residual. This simplified estimate neglects the additional variance introduced by spectrometry background subtraction, the error propagated through the calibration curve used to quantify the mapping pixels, and assumes optimal instrument conditions. Nested within the error estimation of the calibration curve are the peak and background uncertainties of the selected reference points and those of the standard \cite{Marinenko.2010}. Although these latter measurements were acquired for longer times than the mapped pixels, their associated variance only compounds the total propagated uncertainty. Because a full propagation of these sequential uncertainties would only widen the error margin further, the conservative 15\% lower bound is sufficient to demonstrate the validity of the approach. In short, the fact that the model prediction falls within this irreducible error margin confirms that the proposed technique is a robust estimator of the partial equilibrium distribution of interstitially alloyed species.

Readers interested in a microstructure reconstruction for full-field modelling can access the three-dimensional use case in the supplementary materials (see Subsection~\ref{3d-use-case}). Such application might appeal to researchers developing and deploying full-field micromechanical models encompassing spatially resolved chemical heterogeneities \cite{Pinomaa.2020, Horbach.2024, Wu.2025}.

\subsection{Multi-phase, multi-element example}
\begin{figure}[tb]
\makebox[\textwidth][c]{\includegraphics[width=\textwidth]{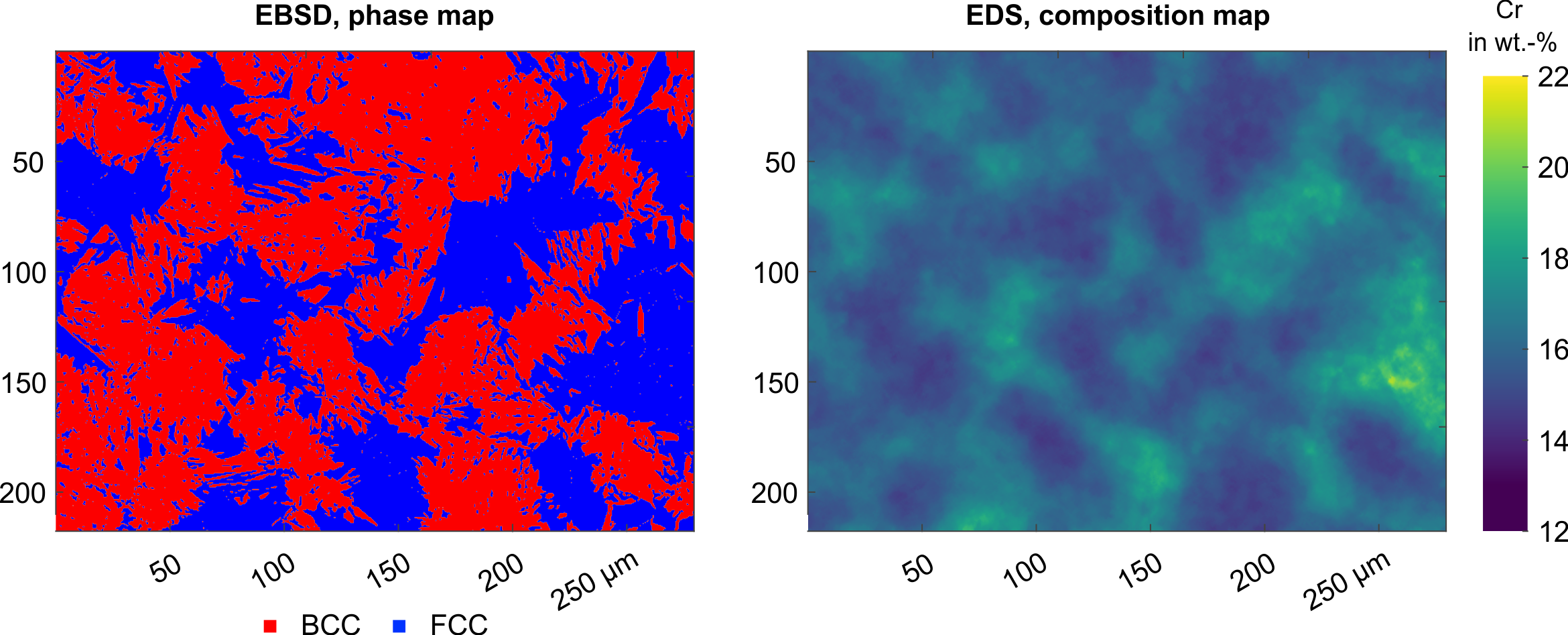}}
\caption{Co-located EBSD and EDS data of the employed material. (a) Phase map. (b) Cr concentration map in wt.\%.}\label{fig:2d-multi-maps}
\end{figure}

Having demonstrated the equivalence of partial equilibrium calculations to short-term diffusion simulations and having validated two-dimensional computations with experimentally obtained composition maps, we finally present in the following the third and final use-case. In it, we show the application of the proposed approach to a carbon and nitrogen alloyed austenitic-bainitic microstructure. The material employed is the commercially available steel \mbox{X30CrMoN15-1} (DIN EN 1.4108) with the chemical composition provided in Table~\ref{14108_composition}. We consider a specific microstructural state generated by austenitization at \SI{1100}{\degreeCelsius} for \SI{20}{\minute}, quenching, and subsequent isothermal holding at \SI{500}{\degreeCelsius} for \SI{24}{\hour}. Figure~\ref{fig:2d-multi-maps} presents the resulting phase distribution, together with the co-located chromium concentration map. Concentration maps for the elements Si, Mn, and Ni were included in the calculations but are omitted here for clarity. One can see in the chromium concentration map a degree of partitioning: austenite regions are enriched with chromium at the expense of the bainite ones.

The microstructural information was acquired using a TESCAN MIRA3 SEM, equipped with Oxford Instruments EBSD and EDS detectors. Operating parameters were: acceleration voltage of \SI{20}{\kilo\volt}, working distance of \SI{17}{\milli\metre}, and specimen tilt angle of \SI{70}{\degree}. The EBSD detector was introduced to a distance of approximately \SI{5}{\milli\meter} from the specimen. Post-processing was kept to a minimum, utilizing only a median filter with a box kernel of side \SI{7}{\pixel}.

From the complete solid solution state in the austenitization step and during the isothermal hold, bainite (indexed as BCC in Figure~\ref{fig:2d-multi-maps}) forms. The scientific and technological interest in this investigation stems from the relatively unexplored bainitic transformation in C+N-alloyed steels. König et al. have shown, for instance, that nitrogen additions accelerate the transformation, attributing this kinetic shift to the precipitation of nanoscale Cr\textsubscript{2}N particles that serve as local nucleation sites and increase the thermodynamic driving force for bainite formation by locally depleting nitrogen \cite{Koenig.2025}. The concurrent redistribution of these fast-diffusers, coupled with localized proneness to build precipitates, continuously alters the driving force for further phase transition. In the context of this work, this material in general, but the considered microstructural state especially, offers the following opportunities:

\begin{enumerate}[label=(\roman*)]
    \item Simultaneous consideration of two interstitial species (C and N).
    \item Definition of distinct spatial regions governed by different host phases.
    \item Evaluation of a metastable state where interstitial partial equilibrium is nonetheless attainable.
\end{enumerate}

Although the binary phase map in Figure~\ref{fig:2d-multi-maps} seems to imply phase-pure austenite and bainitic ferrite regions through its simple bi-color appearance, nano-scale carbides and nitrides, as already mentioned, are known to precipitate at this holding temperature in the considered system. This application therefore presents a multi-scale problem: the observed mesoscale phase and compositional distributions coexist with an underlying precipitate population that remains unresolved at the imaging scale.

Carbides and nitrides are thus not defined as separate regions. At each pixel, in addition to \texttt{FCC\_A1} and \texttt{BCC\_A2} (representing austenite and bainite, respectively), M\textsubscript{2}N (\texttt{HCP\_A3}) and cementite (\texttt{CEMENTITE\_D011}) are included as entered phases. Cementite is widely reported to form at temperatures above \SI{300}{\degreeCelsius} \cite{Berns.2008}, while dichromium nitride has been observed in this temperature range in high-nitrogen steels \cite{Tao.2014}.

Enabling the metastable modelling described in the following is this working hypothesis: At \SI{500}{\degreeCelsius}, the \SI{24}{\hour} isothermal hold provides sufficient time for the highly mobile interstitial elements to redistribute and achieve a state of partial chemical equilibrium, yet it is insufficient to reach equilibrium precipitate phase fractions. As a result, applying the unconstrained partial equilibrium calculation to this system yields one of two extremes: (i) if all precipitate phases are entered at all positions, the precipitation is allowed to proceed to the equilibrium condition, virtually depleting the matrix of C and N, and (ii) if no precipitate phases are entered, the complete solid solution state forced upon the system leads to stark partitioning of C and N to the austenite regions due to the larger octahedral voids in the FCC lattice. Both of these extremes are documented in the Supplementary Materials (\ref{precipitate-equilibrium} and \ref{no-precipitate-equilibrium}). Region-specific aggregates are presented in the columns \emph{Multi-phase}, \emph{Multi-phase, matrix}, and  \emph{Single-phase} of Table~\ref{multi-element-aggregates}, where the former two correspond to the precipitate equilibrium condition and the latter to the calculation without precipitates. Satisfactorily resolving the spatial distribution of these interstitials hinges on the fact that the true microstructural state exists somewhere between the equilibrium precipitation maximum and a complete solid solution.

\begin{table}[tb]
\centering
\footnotesize
\caption{Partitioning of C and N in bainite and austenite regions, region aggregates in wt.\%. All values were obtained through partial equilibrium calculations, with the exception of the metastable aggregates of N, which were used as inputs for the metastable calculation.} \label{multi-element-aggregates}
\small
\begin{threeparttable}
\begin{tabular*}{\textwidth}{@{\extracolsep{\fill}}llccccc}
\toprule
 & Region & \makecell{Multi-phase} & \makecell{Multi-phase\\ matrix} & \makecell{Single-phase} & \makecell{Metastable} & \makecell{Metastable \\ matrix}  \\
\midrule
\multirow{2}{*}{N} & BCC & 0.355 & 0.000 & 0.001 & 0.32\tnote{a} & 0.000 \\
                   & FCC & 0.481 & 0.001 & 0.897 & 0.52\tnote{a} & 0.074 \\
\midrule
\multirow{2}{*}{C} & BCC & 0.382 & 0.000 & 0.046 & 0.29\tnote{b} & 0.001 \\
                   & FCC & 0.260 & 0.001 & 0.671 & 0.36\tnote{b} & 0.047 \\
\bottomrule
\end{tabular*}
  \begin{tablenotes}
    \item[a] Obtained from WDS measurements.
    \item[b] Calculated.
  \end{tablenotes}
\end{threeparttable}
\end{table}

To model this metastable state, we recast its definition as a new optimization problem. An optimization loop wrapping two separate partial equilibrium calculations looks for the interstitial concentrations that best reproduce experimentally measured, region-aggregate concentrations. One calculation handles the precipitation, while the other deals with the solid solution. In the former, therefore, \texttt{HCP\_A3} and \texttt{CEMENTITE\_D011} and the corresponding matrix phase (\texttt{FCC\_A1} or \texttt{BCC\_A2}) are set as entered, whereas in the latter the precipitate phases are suspended. For C and N independently, the optimizer searches for the split of the total available amount of the interstitial species to feed to each calculation. By allocating fewer atoms to the multi-phase calculation for precipitation, we make them available to the matrix-only model, thus approximating---through experimentally obtained data---the true metastable state observed through analytic microscopy.

From the previous paragraph, it is plain that macroscopic chemical composition measurements are not sufficient. To drive the model, region-specific concentrations are required. Column \emph{Metastable} in Table~\ref{multi-element-aggregates} shows the considered region-specific nitrogen contents. These were obtained through additional WDS measurements using the previously mentioned TESCAN MIRA3 SEM at the same operation parameters. The carbon concentrations were, in contrast, not measured and thus taken as optimization variables. Because the characteristic X-ray emission peaks of C and N severely overlap, simultaneous quantification with only one WDS detector is impossible. Without access to a multi-spectrometer microprobe, obtaining both necessitates sequential measurements at distinct locations, increasing experimental burden. 

To summarize: The optimization scheme wrapping the precipitation and solid solution partial equilibrium calculations optimizes how much C and N are available to each model---considering mass conservation---such that the N partitioning agrees with WDS-measured N contents. The choice of this loss function formulation has its grounds on practical terms. As an exercise in measurement efficiency, the calculation predicts C partitioning from nitrogen partitioning. An optimization considering both measured C and N contents (implying a rewrite of the loss function) could naturally also be considered, provided the WDS measurement data were available.

\begin{figure}[p]
\makebox[\textwidth][c]{\includegraphics[width=\textwidth]{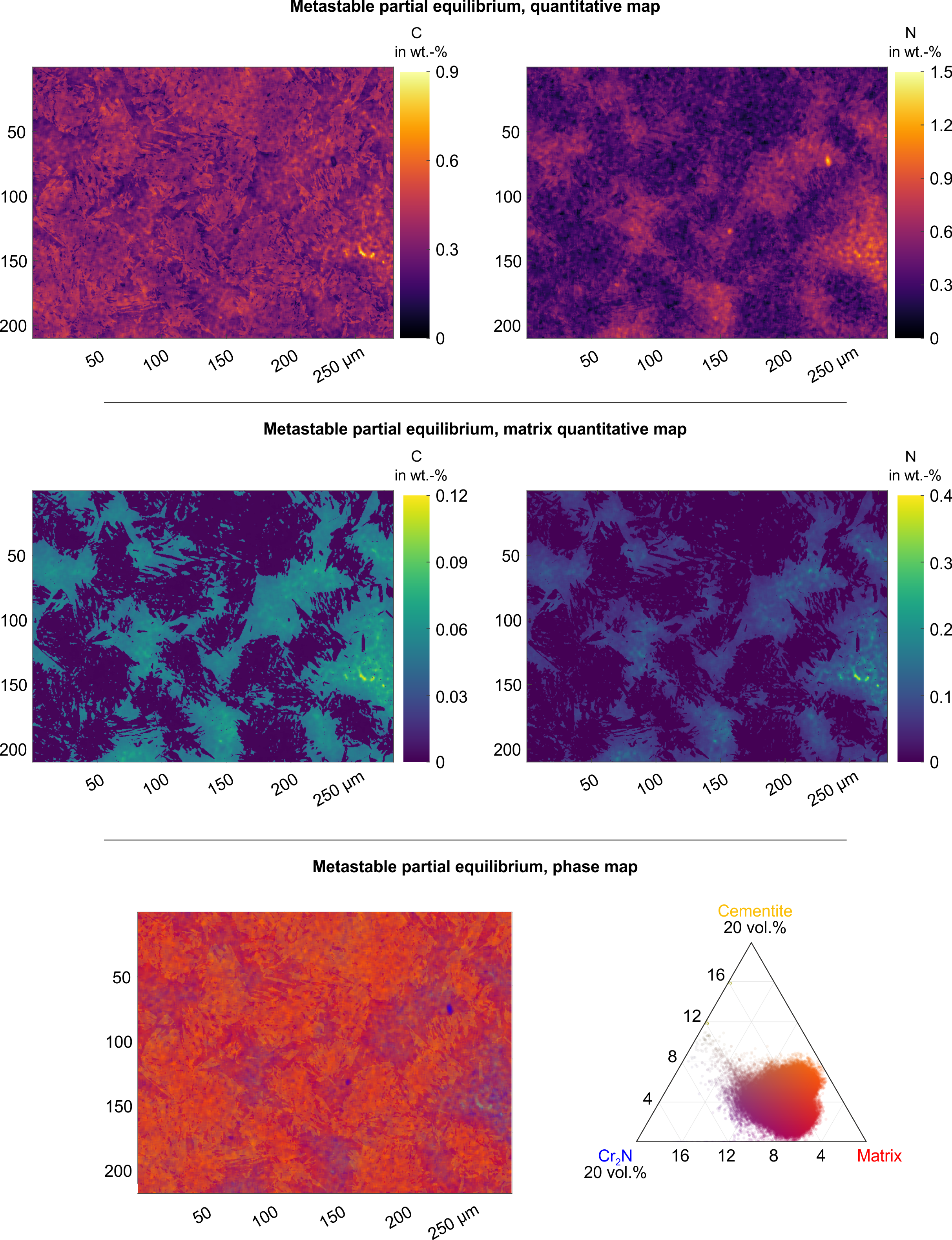}}
\caption{Results of the multi-element, multi-phase metastable partial equilibrium calculation. Top: Distribution of C and N in wt.\%. Middle: Concentration of C and N in the matrix phase in wt.\%. Bottom: Distribution of the entered phases.}\label{fig:2d-multi-metastable-calculation}
\end{figure}

Figure~\ref{fig:2d-multi-metastable-calculation} shows the results of the metastable equilibrium calculation. The top panel presents the element-wise addition of the C and N contents of both models, thus representing their spatial distribution at the mapped mesoscale. The optimization converged on the following splits of the interstitial species to reproduce the experimental measurements. For N, 91.9\% of the total available amount is allocated to the precipitation-active calculation, leaving the remaining 8.1\% to distribute in the plain solid solution model. Carbon exhibits a similar behavior, with 93.5\% directed to the multi-phase calculation. 

Starting with general observations, we point out the partitioning of carbon. The outline of the phase map is clearly visible in the mesoscale elemental map, an indicator of the distinct tendency to form the considered precipitates within the host phase---at least in this equilibrium calculation. At the considered mesoscale, nitrogen is instead driven by the chromium concentration.

In the middle panel, the matrix concentrations show that the precipitates trap significantly less carbon and nitrogen than predicted under equilibrium conditions. Because the model with entered precipitate phases virtually depletes its own matrices from C and N, the optimized \emph{mixture}, i.e., the metastable state, supersaturates the matrix compared to the equilibrium condition. The aggregate metastable matrix concentrations come out to be \SI{0.074}{\wtpercent} N and \SI{0.047}{\wtpercent} C for the austenite region, while the bainite region are virtually depleted of both elements. These values are entered in the last column of Table~\ref{multi-element-aggregates}.

The computed phase fractions are displayed in the bottom panel of Figure~\ref{fig:2d-multi-metastable-calculation}. The calculation predicts an overall higher cementite content in bainite, juxtaposed with predominant dichromium nitride precipitation in Cr-rich, austenite regions. By restricting the interstitial amounts available for precipitation, the mean precipitate volume fraction is reduced from the equilibrium maximum of \SI{9.6}{\volpercent} to approximately \SI{8.9}{\volpercent}, implying that the achieved metastable equilibrium is indeed comparatively close to the partial equilibrium state.

At this point, we would like to reiterate and emphasize the primary objective of both this optimization framework and of the broader work. Here, we pursue the thermodynamically consistent prediction of the spatial distribution of the interstitial solid solution concentrations at the mesoscale. While the model necessarily computes the corresponding metastable precipitate phase fractions to satisfy mass conservation and the experimentally observed partitioning, corroborating the morphology and distribution of these carbides and nitrides falls outside the considered scale and current scope. Ultimately, these predictions show good agreement with previous research on this material \cite{Koenig.2025}.

The framework isolated the excess C and N remaining in solid solution, and importantly, their final spatial distribution. In a system like the one under consideration, these localized interstitial supersaturation states dictate mesoscale austenite stability and precipitation driving forces, to name but a few relevant microstructural properties. Mapping these distributions is therefore a necessary step for deriving property maps discussed in the next and final section.

\section{Summary \& outlook}\label{conclusion}

In this work, we presented an iterative thermodynamic modelling strategy to predict the spatial distribution of interstitially alloyed elements from robust, low-noise compositional maps of heavier substitutional elements. The approach yields a thermodynamically consistent augmentation of quantitative microscopy data by incorporating bulk composition measurements of the fast-diffuser species to drive an optimization scheme. This optimization seeks to establish a state of partial chemical equilibrium for the highly mobile elements.

Following the thermodynamic, mathematical, and algorithmic definition of the problem, the framework was applied to three distinct examples. First, the equivalence of the partial equilibrium approach to short-term diffusion simulations was confirmed against a one-dimensional simulation of a Si-segregation profile in an austenite region. Second, the spatial predictive capability was validated, thus demonstrating that the computed interstitial maps achieve strong quantitative agreement with WDS-obtained nitrogen distributions. Finally, the third example highlighted the flexibility of the methodology. By successfully navigating a multi-phase, multi-element system, the strategy successfully evaluated concurrent precipitation and solid-solution partitioning integrated in a metastable model.

We posit that the described method has the potential to retrieve otherwise inaccessible spatial compositional information, provided the underlying partial chemical equilibrium assumption holds for the investigated system.

As an outlook to illustrate a possible application of the augmented analytic microscopy data, we briefly consider the calculation of property maps.
\subsection{Outlook\# 1: Property maps}

Property maps are derived from the combination of both measured (for the substitutionally alloyed elements) and computed (for the fast-diffusers) composition arrays to assess different mesoscale properties. In the following, we consider two specific thermodynamic properties: phase stability and normalized precipitation driving force. With appropriate physical models, however, a wide array of alternative material descriptors---such as local lattice parameters or thermal and electrical conductivity---could be extracted. Property maps are derived at each coordinate from the local chemical composition, phase distributions, and other relevant quantitative microscopy data such as crystallite orientation \cite{vangenHassend.2020}. For each location, a specific model is then applied to estimate the desired property. In keeping with the use-case described above, the property maps will be calculated on the final matrix compositions delivered by the iterative mesoscale model.

\begin{figure}[tb]
\makebox[\textwidth][c]{\includegraphics[width=.6\textwidth]{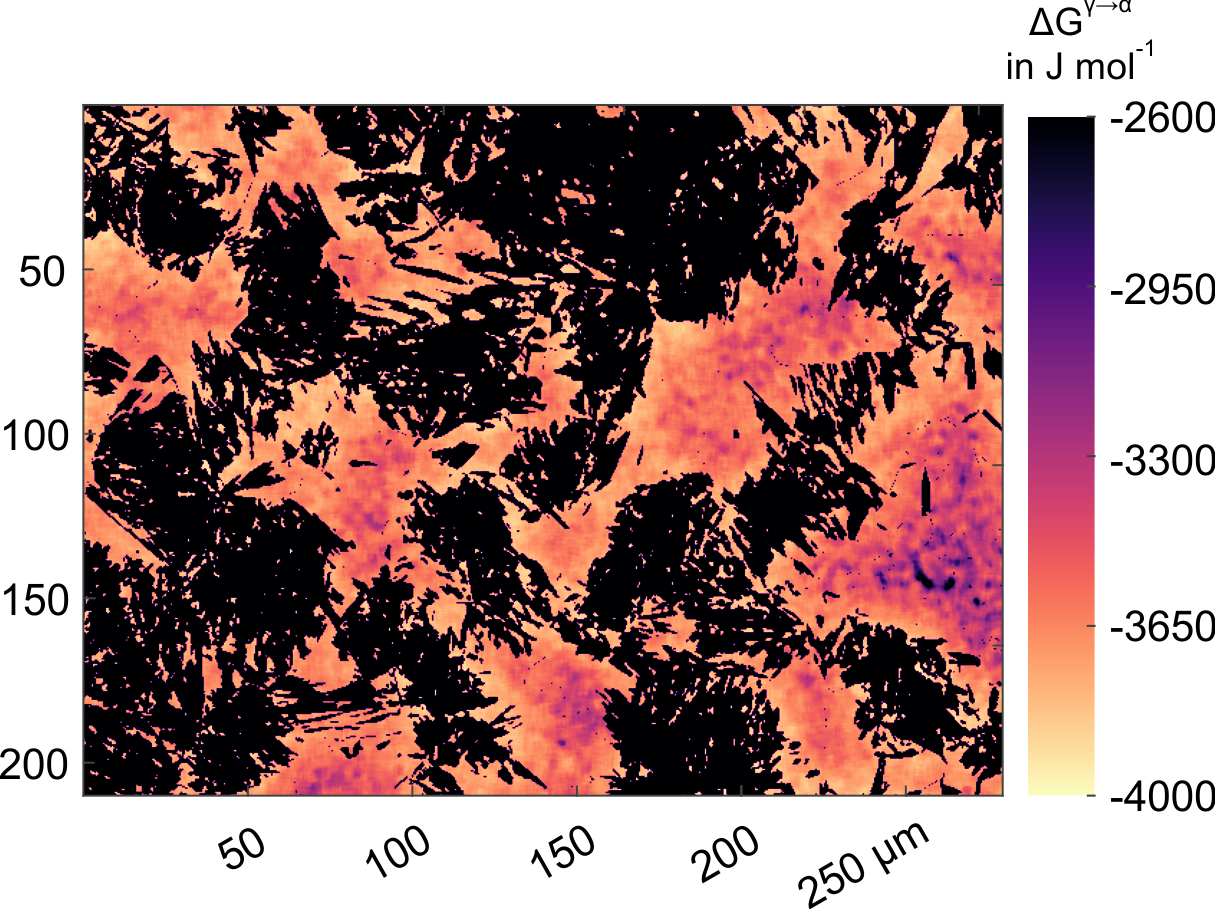}}
\caption{Computed $\Delta G^{\gamma \rightarrow \alpha}$ map. More negative values indicate a higher driving force for the transformation. The BCC regions were excluded from the computation and therefore are blacked out.}\label{fig:outlook_delta-G}
\end{figure}

We first discuss the austenite stability map shown in Figure~\ref{fig:outlook_delta-G}. This map represents the local chemical driving force for ferrite formation, $\Delta G^{\gamma \rightarrow \alpha}$, at a temperature of \SI{300}{\kelvin}. The BCC regions shown in Figure~\ref{fig:2d-multi-maps} were therefore masked out from the calculations, as assessing the austenite stability of these areas is obviously meaningless. Importantly, the regions where the austenitic matrix is locally enriched with C and N exhibit the highest austenite stability. These chemical inhomogeneities stem from the material processing itself, and are, when considering technological applications, to a certain irreducible in nature.  

\begin{figure}[tb]
\makebox[\textwidth][c]{\includegraphics[width=\textwidth]{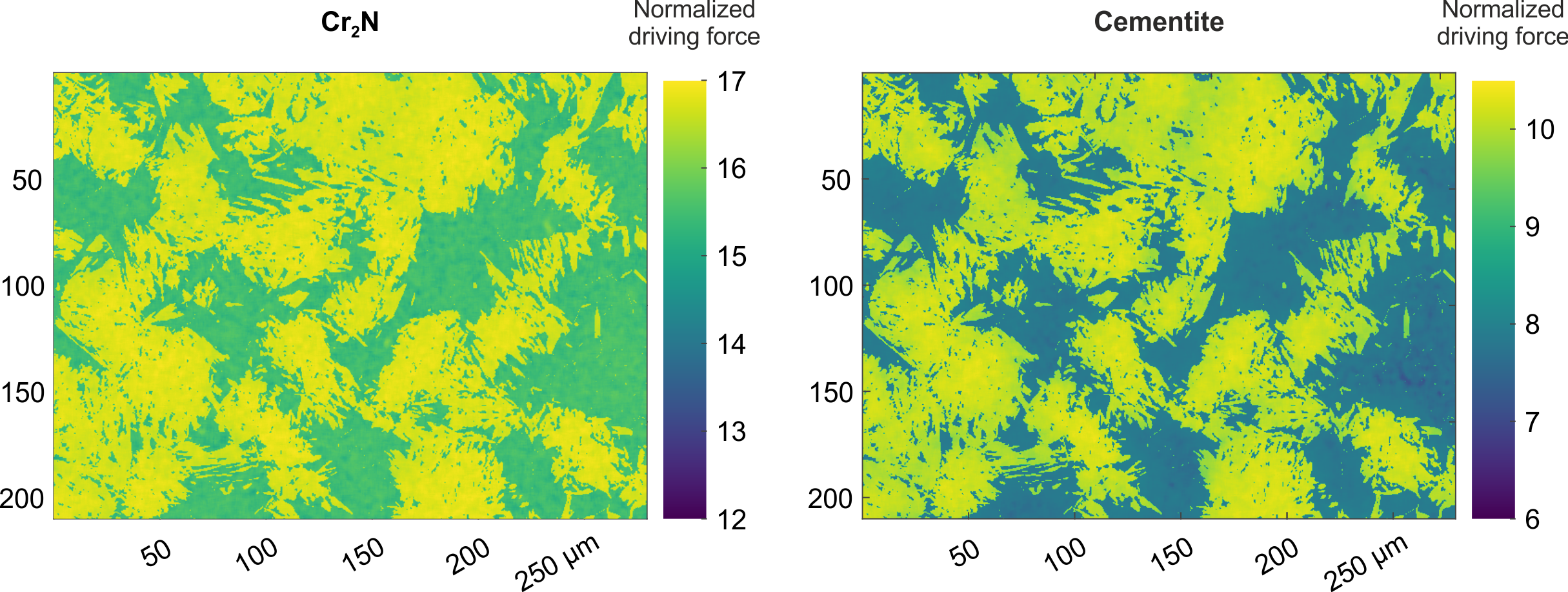}}
\caption{Computed normalized driving force map for the precipitate phases considered in the metastable partial equilibrium calculations. (a) Dichromium nitride. (b) Cementite.}\label{fig:outlook_driving-force}
\end{figure}

The local normalized precipitation driving forces are shown in Figure~\ref{fig:outlook_driving-force}, which is divided into two phase-specific panels. The normalization is performed by dividing the absolute driving force by the thermal energy $RT$, where $R$ is the gas constant and $T$ equals \SI{300}{\kelvin}. The property maps reveal that the precipitate driving forces are higher in the bainitic matrix than in the austenite. The nitrides present overall higher driving forces, owing to the higher nitrogen content in the alloy. Moreover, the calculations indicate that high local Cr-concentrations in the austenite, together with the co-located increase in nitrogen content, result in a highly localized precipitation potential for Cr\textsubscript{2}N in these specific positions.

The spatial variance in these properties cannot be captured by macroscopic bulk measurements or simple X-ray energy-dispersive spectroscopy elemental mapping. The thermodynamically consistent augmentation of the spatially resolved data uncovers localized microstructural features and, subsequently, the true mesoscale material behavior. Ultimately, the combined factors of elemental partitioning, localized precipitate potential, and processing-related inhomogeneities lend distinct phase stabilities to different regions, directly governing the evolution of the microstructure and its final micromechanical properties.

\subsection{Outlook\# 2: Segregation of interstitials to defects}

The framework described in this work is currently unable to model segregation to defects such as grain boundaries (GBs) due to the specific working hypotheses at its core. However, with knowledge of the crystallographic texture from a co-located EBSD-EDS measurement campaign, a solution to this problem takes shape. Based on the easily accessible GB type derived from crystallite misorientation, each GB can, in principle, be assigned an excess Gibbs free energy term \cite{Raabe.2014, Lejcek.2010}. The viability of coupling mesoscale chemical potential equilibrium with spatially resolved defect networks has been recently demonstrated by Traka et al. \cite{Traka.2024}, who successfully utilized EBSD misorientation data to map localized defect densities and compute the concurrent equilibrium between interstitial solid solutions and trapped states. By introducing this localized thermodynamic addition during the chemical potential optimization, our framework would naturally favor interstitial partitioning to the boundaries in accordance with observed segregation trends. We therefore point toward the integration of site-specific crystallographic texture into the employed data stack as an immediate pathway for extending the framework's capabilities.

\subsection{Outlook\# 3: Microstructure reconstruction}

As a final remark, we point out that the approach put forward can be directly used to generate compositional data in the context of Microstructure Characterization and Reconstruction (MCR). MCR modules like MCRpy \cite{Seibert.2022} and Kanapy \cite{Prasad.2019} generate synthetic microstructure data from limited observations. Indeed, both examples can generate three-dimensional microstructures from two-dimensional data. However, to the knowledge of the authors, no MCR software can currently generate compositional maps. A primary challenge is the creation of synthetic data that is consistent with the thermodynamics of the system considered. By this, we mean that explicitly accounting for spatial heterogeneity and the resulting element covariance adds significant complexity: Composition mapping requires the simultaneous generation of interdependent elemental fields. The methodology presented here bridges this gap, offering a robust thermodynamic engine to synthesize spatially resolved, multi-element concentration arrays for advanced MCR applications.

\section*{Acknowledgements}
The authors thank Mr. Philip König for providing the quantiative data employed in the multi-phase, multi-element application example. The authors also express their gratitude for the funding of the vacuum induction melting inert gas atomization system by the DFG with the grant number 470572383.

\bibliographystyle{elsarticle-num} 
\bibliography{Bibliography}
\appendix
\newpage
\setcounter{section}{0}
\setcounter{page}{1}
\setcounter{figure}{0}
\setcounter{table}{0}
\setcounter{equation}{0}
\renewcommand{\thesection}{S\arabic{section}}
\renewcommand{\thepage}{s\arabic{page}}
\renewcommand{\thetable}{S\arabic{section}}
\renewcommand{\thefigure}{S\arabic{section}}

\setcounter{affn}{0}
\resetTitleCounters

\makeatletter
\let\@title\@empty
\makeatother

\title{Supplementary Data for: Iterative Thermodynamic Augmentation of Spatially Resolved Analytic Microscopy for Fast-Diffusing Solutes}

\makeatletter
\renewenvironment{abstract}{\global\setbox\absbox=\vbox\bgroup
  \hsize=\textwidth\def\baselinestretch{1}%
  \noindent\unskip\textbf{Contents}
 \par\medskip\noindent\unskip}
 {\egroup}


\startlist{toc}
\begin{abstract}
\footnotesize
\vspace{-48pt}
\printlist{toc}{}{\section*{}}
\end{abstract}
\maketitle
\section*{}

\section{Chemical composition of the X2CrNi18-9 powder} \label{304L_composition}

\begin{table}[H]
\centering
\setlength{\tabcolsep}{3pt}
\footnotesize
\caption{Chemical composition of the X2CrNi18-9 powder (wt.\%).}
\begin{tabular*}{\textwidth}{@{\extracolsep{\fill}}ccccccccc}
\toprule
C & Si & Mn & P+S & Cr & Ni & N & O & Fe \\
\midrule
$<0.03$ & 
\makecell{0.85 \\ $\pm$ 0.15} & 
\makecell{1.3 \\ $\pm$ 0.16} & 
$<0.03$ & 
\makecell{18.92 \\ $\pm$ 0.27} & 
\makecell{9.11 \\ $\pm$ 0.72} & 
\makecell{0.084 \\ $\pm$ 0.01} & 
\makecell{0.03 \\ $\pm$ 0.005} & 
bal. \\
\bottomrule
\end{tabular*}
\end{table}

\section{Further results of the diffusion simulations}

\begin{figure}[H]
\makebox[\textwidth][c]{\includegraphics[width=\textwidth]{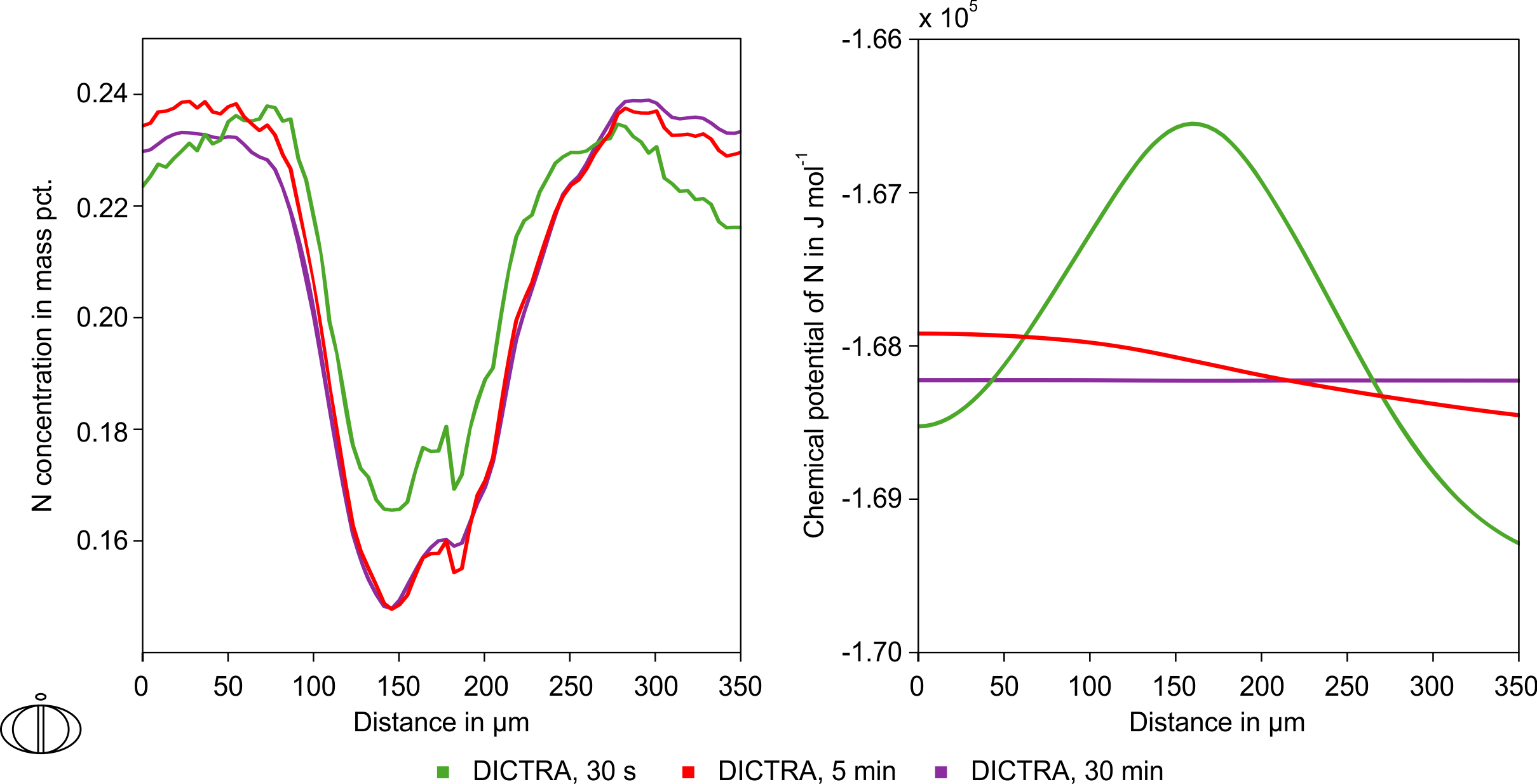}}
\caption{Results of the diffusion simulations at different simulation times.}\label{fig:1d-example-dictra}
\end{figure}

\section{Chemical composition of the Fe20Cr powder} \label{fe20cr_composition}

\begin{table}[H]
\centering
\setlength{\tabcolsep}{3pt}
\footnotesize
\caption{Chemical composition of the Fe20Cr powder (wt.\%).}

\small
\begin{tabular*}{\textwidth}{@{\extracolsep{\fill}}cccccc}
\toprule
 Si & P+S & Cr & N & O & Fe \\
\midrule
\makecell{0.01 \\ $\pm$ 0.001} & 
$<0.03$ & 
\makecell{19.59 \\ $\pm$ 0.04} & 
\makecell{0.014 \\ $\pm$ 0.000} & 
\makecell{0.04 \\ $\pm$ 0.002} & 
bal. \\
\bottomrule
\end{tabular*}
\end{table}

\section{Additional concentration maps of the Fe20Cr-Si\textsubscript{3}N\textsubscript{4} mixture} \label{fe20cr_maps}

\begin{figure}[H]
\makebox[\textwidth][c]{\includegraphics[width=\textwidth]{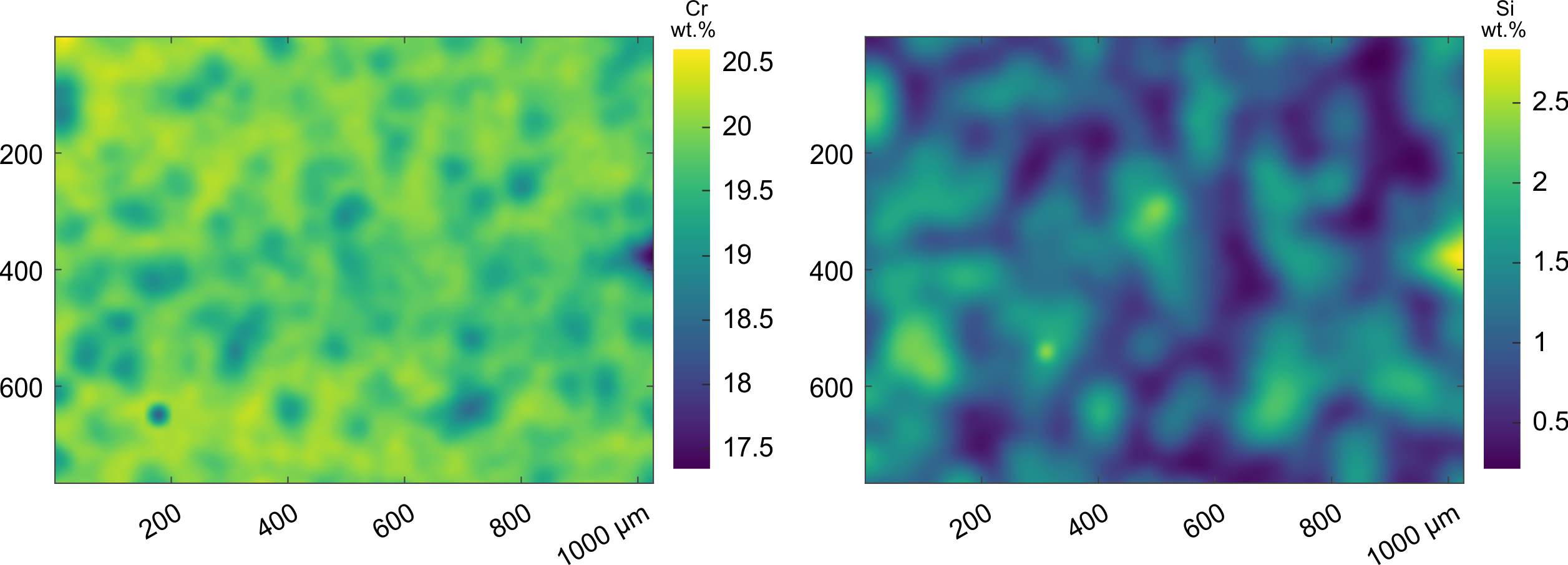}}
\caption{Measured chromium and silicon concentrations. These composition maps constitute the basis for the partial equilibrium distribution of nitrogen presented in Figure \ref{fig:2d-example-maps}}
\end{figure}

\section{Definition of the normalized first-order variogram} \label{normalized-variogram}
Assuming first-order stationarity, we employed the fast-Fourier implementation of the method of moments estimation of the variogram proposed my Marcotte \cite{Marcotte.1996}:

\begin{equation}
    2\gamma = \frac{f^2 \star I_{f} + I_{f} \star f^2 -2(f \star f)}{I_{f} \star I_{f}},
\end{equation}

where $f$ is the composition map, $I_{f}$ is an index matrix, and $\star$ represents the correlation operation. The index matrix contains the number one at each map valid position. Finally, the division operation is performed element-wise.

The standard variogram, $2\gamma(h)$, was converted in a normalized first-order variogram, $2\hat{\gamma_1}(h)$ \cite{Benito.2023}:

\begin{equation}
    2\hat{\gamma_1}(h) = \sqrt{\frac{\gamma(h)}{\text{var}(f)}},
\end{equation}

where $h$ is the distance vector at which the mean absolute difference is computed. $\text{var}(f)$ is the variance of the composition map, serving as a normalizing factor.

\section{Definition of the cross-correlogram} \label{cross-correlogram}

Employing Marcotte's implementation \cite{Marcotte.1996}, we computed the cross-covariogram as follows:

\begin{equation}
    C = \frac{f \star g}{I_{f} \star I_{g}} - \frac{(f \star I_{g}) (I_{f} \star g)}{(I_{f} \star I_{g})^2},
\end{equation}

where the multiplications and divisions between matrices are performed element-wise. $f$ and $g$ are composition maps, whereas $I_{f}$ and $I_{g}$ are the corresponding index matrices.

The cross-covariogram was then converted into a cross-correlogram by:

\begin{equation}
    \rho(h) = \frac{ C(h)}{\sqrt{\text{var}(f)\text{var}(g)}},
\end{equation}

where, again, $\text{var}(\cdot)$ represents the map variance.

\newpage

\section{Additional correlation analysis} \label{additional-correlation}

\begin{figure}[H]
\makebox[\textwidth][c]{\includegraphics[width=0.7\textwidth]{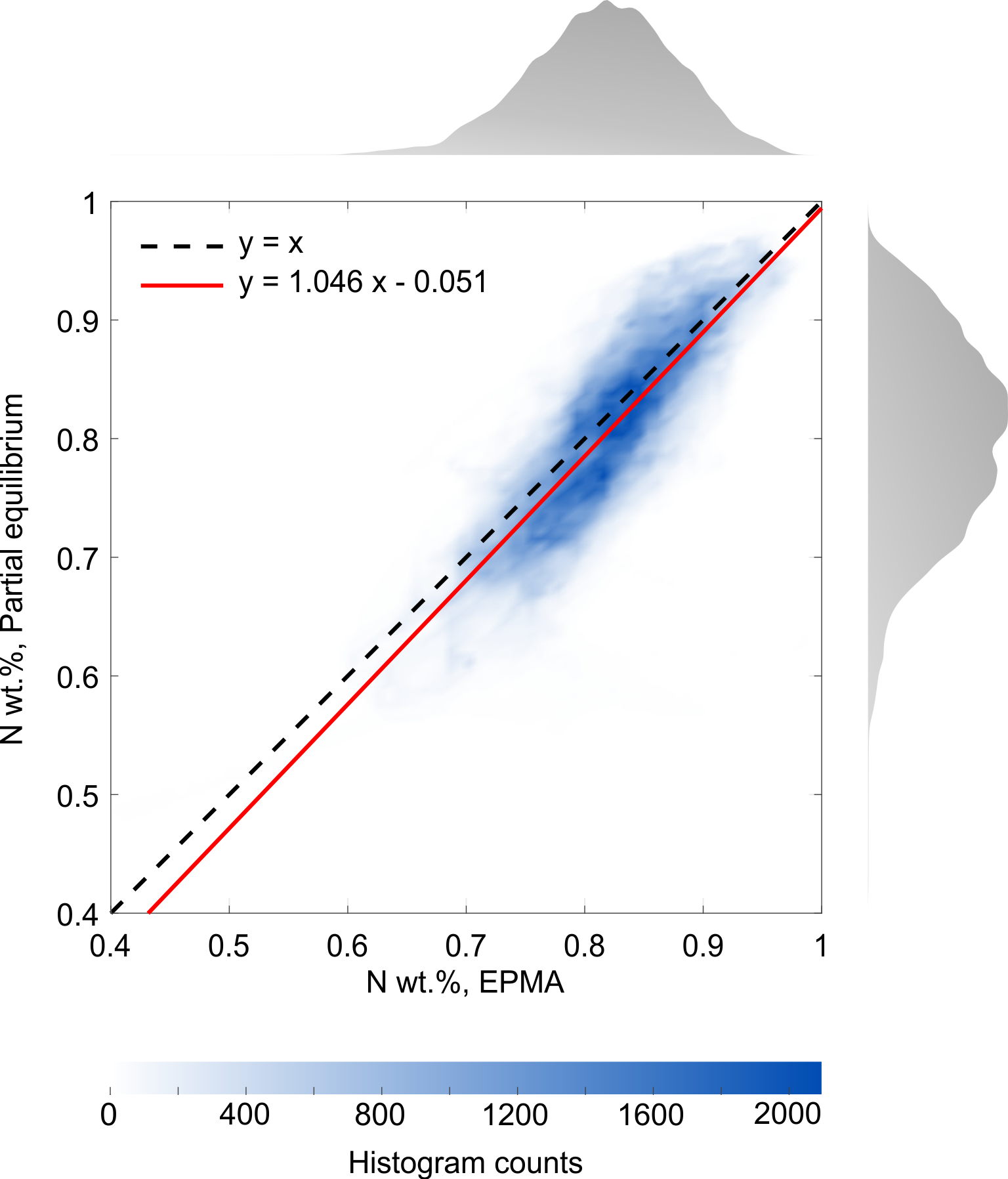}}
\caption{Bivariate histogram comparing the N distribution values obtained through EPMA and the partial equilibrium calculation. The marginal distributions previously shown in Figure \ref{fig:2d-example-statistics} a) are plotted outside the Cartesian axes. The regression analysis on this reliability plot returns an almost perfect $y = x$, albeit with some scatter. The model achieves an adjusted coefficient of determination $R_{adj}^2$ of 0.716. As explained in the discussion in the main manuscript, while a perfect correlation is theoretically possible, due to shot noise this is not practically observable. To put it differently, no deterministic model could (or should) return an $R_{adj}^2$ of unity in this kind of application.} 
\end{figure}

\section{Three-dimensional example} \label{3d-use-case}
This example briefly demonstrates the application of the presented approach to three-dimensional data. As stated in the main manuscript, this kind of application is of relevance when dealing with full-field material models that require spatial compositional information \cite{Pinomaa.2020, Horbach.2024, Wu.2025}. Three-dimensional concentration maps might be acquired through serial sectioning and imaging or synthetically generated through microstructural reconstruction methods based on generative models or descriptor-based stochastic reconstruction algorithms.

The extension of the partial equilibrium calculation described here is straightforward: Optimizing the chemical potentials of the selected interstitial species indeed results in their spatial distribution. However, this is done \emph{globally}, meaning that the actual order of the input array containing the chemical composition of the substitutional sublattice or its shape is irrelevant, as long as the result can be reconstructed after the calculation. In other words, the three-dimensional substitutional data can be reshaped into a two-dimensional array, which the thermodynamic augmentation algorithm then processes to produce the corresponding interstitial maps. A final step is then required to recover the original three-dimensional geometry of the microstructure from the \emph{flattened} array.

\begin{figure}[p]
\makebox[\textwidth][c]{\includegraphics[width=\textwidth]{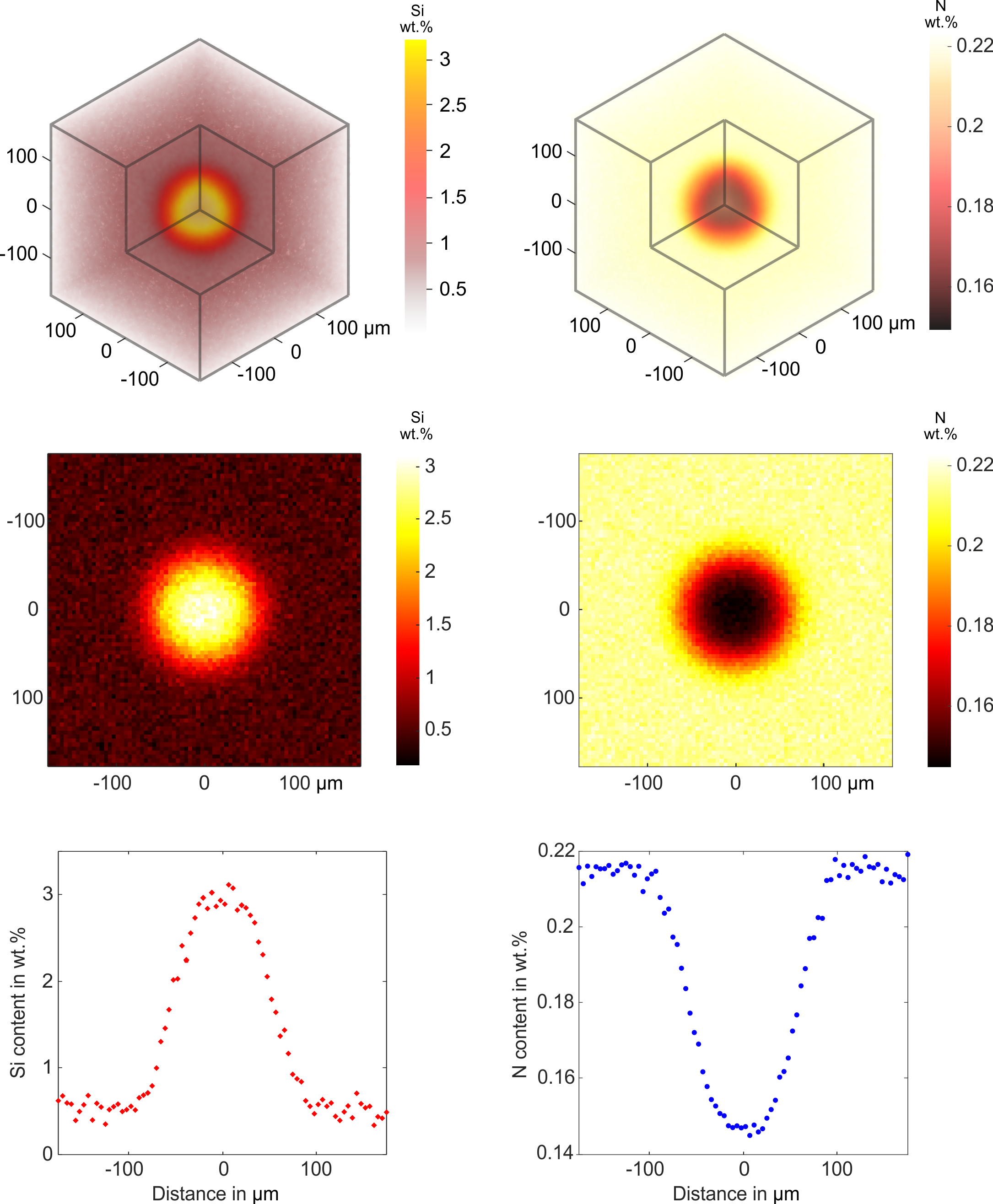}}
\caption{Three-dimensional representation of the computed partial equilibrium. The left column displays the input silicon distribution from a voxel-based reconstruction, while the right column shows the resulting thermodynamically computed nitrogen concentration. The panels display an isometric cutaway view (top row), a two-dimensional mid-plane slice (middle row), and a corresponding line profile across the center of the domain (bottom row).} \label{3d-example}
\end{figure}

In this example, we extended the data presented in Section~\ref{one-dimensional-example} to create a voxel-based reconstruction of a prior silicon nitride particle in the powder mixture. Figure~\ref{3d-example} shows the silicon distribution in the left column, while the right column presents the computed nitrogen concentration. The chromium concentration was also generated and employed in the partial equilibrium calculations, but is omitted here for brevity.

Starting from the top, the three-dimensional data is shown in an isometric view. One-eighth of the cubic domain was removed to clearly show the silicon and nitrogen gradients resulting from the powder mixing alloying strategy (cf. Section~\ref{one-dimensional-example}). The second row presents the middle slice parallel to the base plane, as seen from the top. The third row is a one-dimensional line profile extracted from the center of the cube.

Comparing the results of Figure~\ref{3d-example} with those presented in Figure~\ref{fig:1d-example}, one can see that these are equivalent, the only critical difference being the dimensionality of the output.

\section{Chemical composition of the X30CrMoN15-1 specimen} \label{14108_composition}

\begin{table}[H]
\centering
\setlength{\tabcolsep}{3pt}
\footnotesize
\caption{Chemical composition of the X30CrMoN15-1 specimen (wt.\%).}
\small
\begin{tabular*}{\textwidth}{@{\extracolsep{\fill}}ccccccccc}
\toprule
C & Si & Mn & P+S & Cr & Ni & Mo & N & Fe \\
\midrule
\makecell{0.33 \\ $\pm$ 0.02} & 
\makecell{0.59 \\ $\pm$ 0.03} & 
\makecell{0.43 \\ $\pm$ 0.03} & 
\makecell{$<0.04$ \\ $\pm$ 0.0002} &
\makecell{15.55 \\ $\pm$ 0.05} &
\makecell{0.25 \\ $\pm$ 0.02} &
\makecell{0.95 \\ $\pm$ 0.02} &
\makecell{0.41 \\ $\pm$ 0.01} &
bal. \\
\bottomrule
\end{tabular*}
\end{table}

\section{Multi-element partial equilibrium with precipitates} \label{precipitate-equilibrium}

\begin{figure}[H]
\makebox[\textwidth][c]{\includegraphics[width=\textwidth]{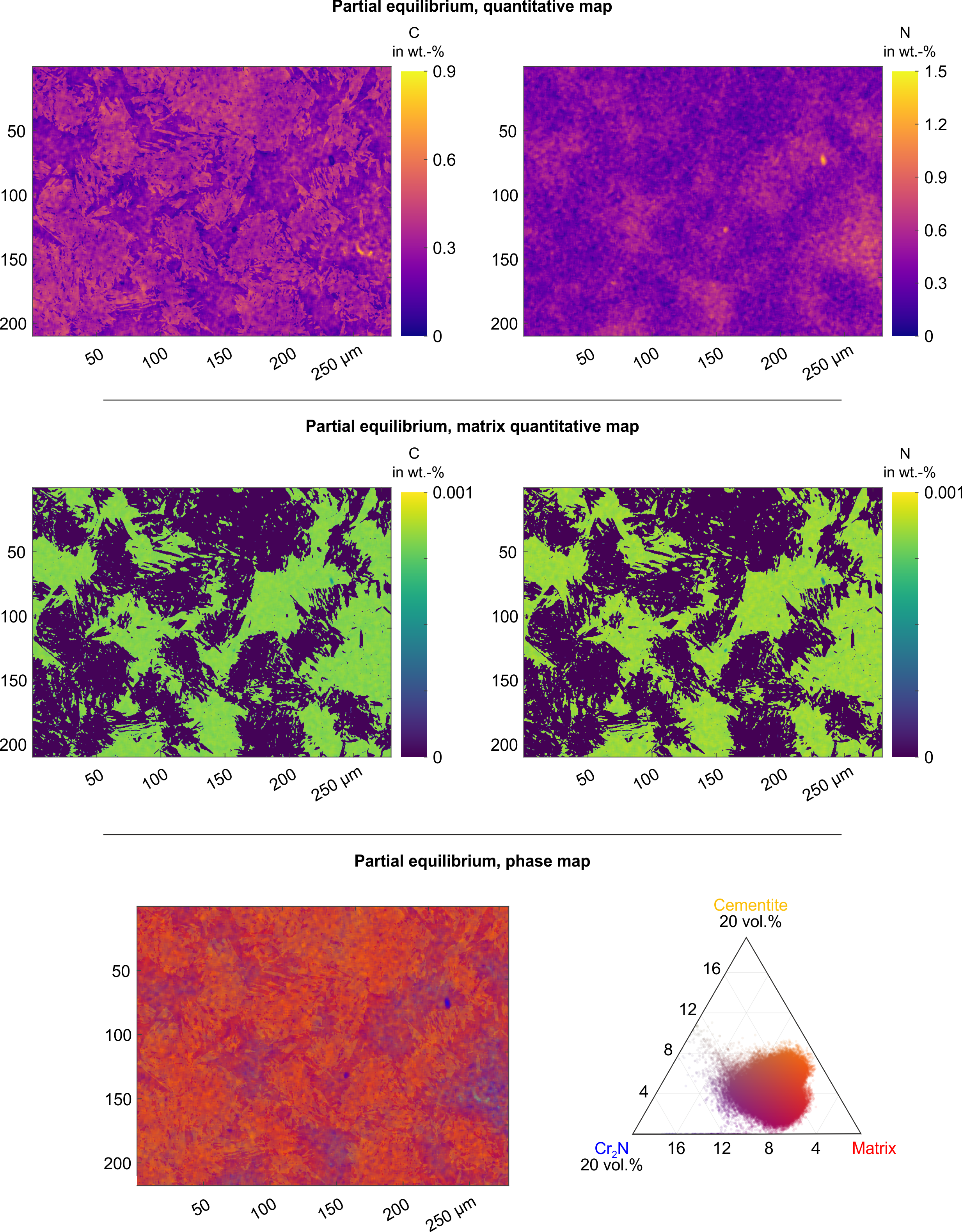}}
\caption{Results of the multi-element, multi-phase partial equilibrium calculation. Top: Distribution of C and N in wt.\%. Middle: Concentration of C and N in the matrix phase in wt.\%. Bottom: Distribution of the entered phases.}\label{fig:2d-multi-first-calculation}
\end{figure}

\section{Multi-element partial equilibrium without precipitates} \label{no-precipitate-equilibrium}

\begin{figure}[H]
\makebox[\textwidth][c]{\includegraphics[width=\textwidth]{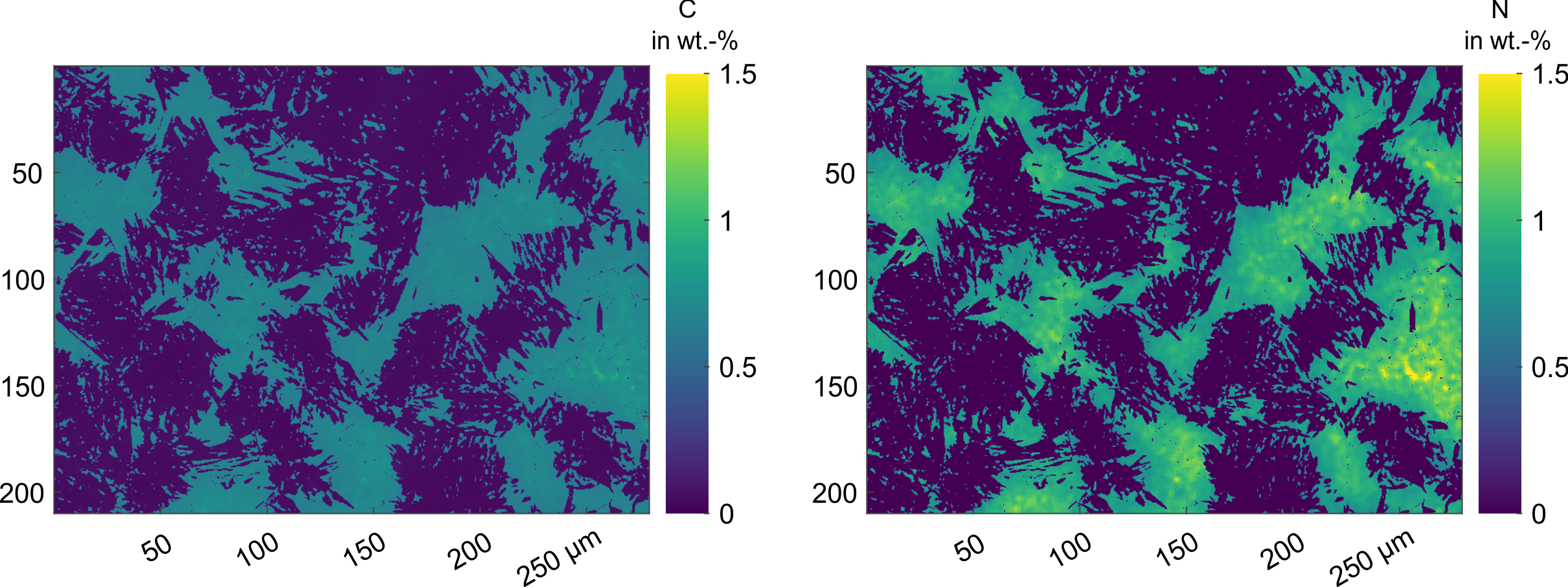}}
\caption{Partial equilibrium distribution of C and N with only the matrix phases entered: \texttt{FCC\_A1} and \texttt{BCC\_A2}. Both fast-diffusers are heavily partitioned, with nitrogen showing virtually no concentration in solid solution in the cubic-centered regions.}
\end{figure}

\end{document}